\documentclass[
aps,
prl,
preprint,
  superscriptaddress,
]{revtex4-2}

\usepackage[utf8]{inputenc}
\usepackage{graphicx}
\usepackage{xcolor}
\usepackage{dcolumn}
\usepackage{bm}
\usepackage[all]{nowidow}
\usepackage{siunitx}[=v2]
\usepackage{ulem}
\usepackage{amsmath}
\usepackage[english]{babel}

\begin{document}
\title{Correlation-induced phase shifts and time delays in resonance enhanced high harmonic generation from Cr$^{+}$}

\author{Yoad Aharon}
\author{Adi Pick}
\author{Amir Hen}
\author{Gilad Marcus}
\email{gilad.marcus@mail.huji.ac.il}
\affiliation{Institute of Applied Physics, Hebrew University of Jerusalem, Jerusalem 9190401, Israel}

\author{Ofer Neufeld}
\email{ofern@technion.ac.il}
\affiliation{Technion Israel Institute of Technology, Faculty of Chemistry, Haifa 3200003, Israel}

\date{\today}

\begin{abstract}
We investigate resonance-enhanced high harmonic generation (rHHG) in Cr$^{+}$ by comparing a 1D shape-resonant model, time-dependent density functional theory (TDDFT), and independent particle approximation (IPA) simulations. Previous studies linked rHHG to the $3p \rightarrow 3d$ giant resonance, suggesting a modified four-step model where recolliding electrons are first captured in the autoionization state before recombining into the ground state, potentially leading to an emission delay associated with the resonance lifetime. While both the 1D-model and TDDFT reproduce experimental spectra, TDDFT reveals that rHHG counterintuitively originates from spin-down $3p$ states, while spin-up $3d$ electrons negligibly contribute. The IPA fails to reproduce rHHG, highlighting the significance of electron correlations. Furthermore, TDDFT revealed a correlation-induced $\sim$480 attosecond time delay, accompanied by a strong phase shift across the resonance, potentially explaining earlier RABBIT measurements. Our work sheds light on long-standing open questions in rHHG and should advance novel ultrafast spectroscopies of electron correlations and resonances. 
\end{abstract}

\maketitle  

High-harmonic generation (HHG) is a fundamental process in strong-field physics, enabling generation of coherent extreme ultraviolet (XUV) and soft X-ray radiation\cite{hadrich_exploring_2015, chang_generation_1997, spielmann_generation_1997, li_attosecond_2020, marcus_subfemtosecond_2012}. It presents a powerful tool for probing ultrafast electron dynamics and structural properties of matter across phases \cite{shafir_resolving_2012, smirnova_high_2009,veisz_generation_2013,bergues_tabletop_2018, sobolev_terawatt-level_2024,mondal_high-harmonic_2023,deng_laser-induced_2020, alcala_high-harmonic_2022, neufeld_probing_2020, neufeld_ultrasensitive_2019, hamer_characterizing_2022, freudenstein_attosecond_2022, neufeld_probing_2022, heide_probing_2022, schmid_tunable_2021, zhang_high-harmonic_2024}. In gases, HHG is well described by the three-step model \cite{krause_high-order_1992, corkum_plasma_1993} whereby an electron: (i) tunnel ionizes, (ii) is accelerated by the laser, and (iii) recombines with the parent ion, emitting a photon. The coherence of the electronic wave function throughout this process is key towards utilizing HHG for ultrafast spectroscopy, as has been demonstrated for charge migration\cite{hamer_characterizing_2022, he_filming_2022}, molecular symmetry and orientation \cite{baykusheva_bicircular_2016, frumker_oriented_2012, neufeld_detecting_2022, saito_observation_2017, neufeld_floquet_2019,PhysRevA.97.023408, PhysRevA.69.031804, torres_probing_2007, madsen_theoretical_2007}, and more.

One known gas-phase system that deviates from this three-step model picture is transition-metal ions. Experimentally, HHG from laser-ablated plasma is known to lead to a significant resonant enhancement of a single harmonic order \cite{suzuki_anomalous_2006, ganeev_generation_2009, rosenthal_discriminating_2015}. Resonant HHG (rHHG) has been established to originate at the single atom level, with macroscopic effects playing a negligible role \cite{rosenthal_discriminating_2015}. The energy of the enhanced harmonic photon aligns with a transition to an autoionization state (AIS) in the cation, related to the $3p\rightarrow3d$ giant resonance. Briefly, this autoionization transition involves the excitation of a $3p$ core electron to an unoccupied $3d$ subshell \cite{costello_3p_1991,osawa_photoion-yield_2012,frolov_potential_2010}, 
after which the resulting highly excited state (e.g., $3p^53d^{n+1}$) is unstable and decays via autoionization, typically filling the $3p$ hole and emitting an outer electron (e.g., $3p^63d^{n-1} + \epsilon l$).
Resonances in photoionization spectra and resonant photoionization have been the subject of extensive study. A complete topical review is provided in \cite{kheifets_resonant_2025}. 
For Cr$^+$, the ground state configuration is $3p^63d^5(^6S)$ in LS coupling notation, corresponding to a half-filled $3d$ sub-shell with five spin-up electrons. This configuration breaks symmetry between spin-up/down electrons due to an exchange interaction.  The dominant excitation is still the $3p\rightarrow3d$ transition, i.e. Cr$^+$ $3p^63d^5(^6S) + \gamma$ → Cr$^+$ $3p^5(^2P)3d^6(^5D)(^6P)$ followed by autoionization into Cr$^{2+}~3p^63d^4(^5D)+\epsilon l$  with $l=3$ being the dominant channel \cite{costello_3p_1991,frolov_potential_2010, osawa_photoion-yield_2012}.

To account for enhancement in rHHG in presence of multi-photon transitions, an extended semi-classical single-active electron (SAE) 4-step model was suggested \cite{strelkov_role_2010}.
It replaces the recombination step of the three-step model with electron capture in the AIS, followed by a transition to the ground state and photon emission. 

Alternatively, one can describe rHHG fully numerically with methods falling into two categories: (i) SAE models where the AIS is approximated by a resonance in a barrier-well potential, forming a metastable state (MS) \cite{strelkov_role_2010, strelkov_high-order_2014, tudorovskaya_high-order_2011, ganeev_isolated_2012}. These models qualitatively reproduces the measured enhancement, but do so by synthetically fitting the transition energies. It remains unclear to what extent the SAE approximation captures the full dynamics of rHHG and if any time-delay or phase shift are inflicted on the resonant photon.
Approach (ii) is to perform \textit{ab-initio} simulations of rHHG using either time-dependent density functional theory (TDDFT) \cite{romanov_simulation_2024, romanov_study_2021}, or multi-configuration time-dependent approaches \cite{wahyutama_time-dependent_2019, tikhomirov_high-harmonic_2017}. These schemes lead to spectra in agreement with experiments without the need of external parameters, but are more difficult to analyze and extract the physical mechanism. Still, with \textit{ab-initio} schemes it was established that in some systems correlations play a dominant role in enabling resonant emission \cite{romanov_simulation_2024}, and that deeper orbitals often play a crucial role in dynamics \cite{wahyutama_time-dependent_2019}. What remains unclear is how these effects transpire, and how they depend on the electronic spin in magnetic ions such as Cr${^+}$. In particular, the phase and time delay of rHHG has not yet been explored with multi-electron theories (only with model semi-analytic approaches\cite{strelkov_high-order_2014}).

We perform here an extensive theoretical analysis of rHHG from magnetic Cr$^{+}$ employing three complementary levels of theory: (i) a shape-resonant SAE model, (ii) \textit{ab-initio} spin-TDDFT simulations, and (iii) simulations on par with the spin-TDDFT but where electron correlations are artificially frozen, denoted as the independent particle approximation (IPA). This comparison allows pin-pointing the mechanism of rHHG - a low-lying $3d$ electron is tunnel ionized and gains high kinetic energy in the laser field. When it re-collides with the parent ion, it promotes a deep-level $3p$ electron via electron-electron interactions into empty $d$-bands, forming the autoionization state, from which an enhanced rHHG photon is emitted. The $3p$ energy level is situated too deep to typically contribute to HHG if electronic interactions are not invoked. We show that this transition, which does not occur in uncorrelated theories, inflicts a time-delay and phase shift onto the rHHG emission, suggesting correlated corrections to the 4-step model. Importantly, we find that this transition is only facilitated by spin-down electrons that can pass through the empty $d$-bands. Our results shed light onto fundamental rHHG physics and could lead to HHG-spectroscopy of attosecond correlated multi-electron systems. 

\begin{figure}[htbp]
  \centering
  \includegraphics[width=0.9\textwidth, trim={0 19.5cm 2.2cm 0}]{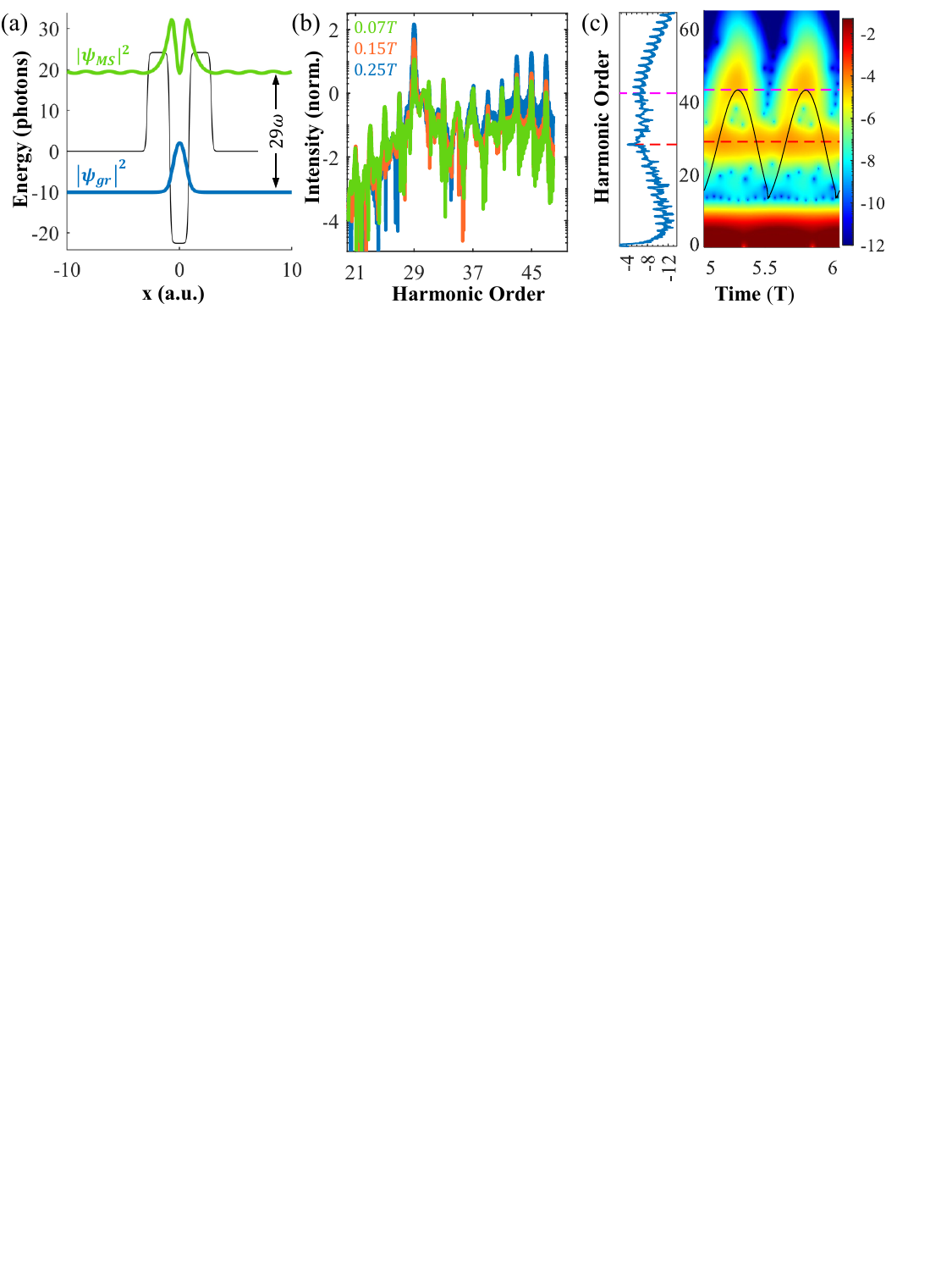}
  \caption{\textbf{rHHG from the 1D model. (a)} Potential supporting a shape resonance (MS, black). The ground state (blue) and MS (green) densities are plotted in their respective eigen-energies. The vertical axis (energy) is normalized to the driving photon energy. \textbf{(b)} HHG spectra for different barrier widths/MS lifetimes. Curves are normalized to the 27th harmonic. The enhancement coefficient increases with the MS lifetime. \textbf{(c)} (Right) Gabor transform of HHG emission for an MS lifetime of 0.17T (T is the period of the driving laser). Red, purple, and black curves indicate the resonant harmonic energy, the semiclassical cutoff energy, and semiclassical emission times (trajectories), respectively.  (Left) Corresponding spectrum. The enhancement of the resonant harmonic H29 is evident in both the spectrum and the Gabor transform. The harmonic emission times are in strong agreement with the semiclassical model.}
  \label{fig:1d_fig}
\end{figure}

Let us briefly describe our numerical approach and examined set-up. First, we explore HHG and corresponding time-frequency plots from a 1D SAE model that includes a shape resonance matching withe the experimental resonant transition (all numerical details are delegated to the Appendix). The results in Fig. \ref{fig:1d_fig} show a good agreement with experimental rHHG spectra from Cr$^{+}$ \cite{rosenthal_discriminating_2015, ganeev_resonance-affected_2021}. We examine the behavior of resonant H29 for different resonance lifetimes (Fig. \ref{fig:1d_fig} b), especially quantifying its enhancement factor, defined as $\frac{I_{29}}{I_{27}}$. In doing so, we hope to resolve whether a re-collision picture is valid in rHHG in the SAE case. While the overall harmonic yield decreases with increasing lifetime (see SI), the enhancement factor increases. This is consistent with a recollision-based picture, since the increase in lifetime is achieved by widening the barrier, thereby decreasing the tunneling rate. To further investigate this interpretation, we gradually extended the complex absorbing potential (CAP) inwards from the box boundary toward its center. As the CAP approached the semiclassical long trajectories, these progressively disappeared from the time-frequency (Fig. S2), while short trajectories were unaffected.
By extending the CAP inward the harmonic yields reduce, including H29, further supporting the recollision mechanism for rHHG.

Time-frequency analysis for an exemplary case is shown in Fig. \ref{fig:1d_fig}(c). The semiclassical trajectories match the quantum SAE simulations with high accuracy --- there is no distinguishable time delay for the resonant H29. We note that according to the 4-step model, H29 is emitted after the electron is captured in the AIS, therefore some delay is anticipated, though we could not resolve it in our conditions.

 Next, we employ an \textit{ab-initio} TDDFT scheme (for details see Appendix). Figure \ref{fig:TDFT_gabor} shows HHG spectra and time-frequency analysis under similar conditions to Fig. \ref{fig:1d_fig} separated to spin-up/down channels (Fig. \ref{fig:TDFT_gabor}(c,d)), and the total spin-summed response (Fig. \ref{fig:TDFT_gabor}(b)). While both channels contribute to the resonance emission, the spin-down channel response is about an order of magnitude larger than the spin-up, dominating rHHG. This is counter-intuitive if considering the typical physical mechanism of atomic HHG and the energy level diagram in Fig. \ref{fig:TDFT_gabor}(b) - the spin-up channel includes initially occupied $3d$ states that have a much lower ionization potential compared to the $3p$ levels. \textit{A-priori}, this should make their tunnel ionization and recombination exponentially more efficient. However, the fact that $3p$ spin-down states dominate the resonant response indicates that they become intrinsically linked to the AIS through interaction with the laser. In these calculations, the long trajectories are slightly suppressed compared to the 1D case, as is common in \textit{ab-initio} theory\cite{neufeld_bench}, due to the 3D nature of the simulation.

Another surprising feature can be seen in the time-frequency plots in Fig. \ref{fig:TDFT_gabor}.
First, we notice that the HHG emission time in the spin-up channel largely follows the semi-classical trajectories with ionization potential matching the $3d$ levels. This is expected since the $3d$ spin-up sub-shell is initially occupied and its ionization energy is much lower than the ionization energy of the $3p$ electrons.
The most striking feature is that the HHG emission from the spin-down channel also largely follows the classical trajectories initiated from $3d$ electrons, even though the $3d$  states are not initially occupied by spin-down electrons. This is a direct indication that rHHG induces a $3p\rightarrow3d$ transition, while this channel is blocked for the $3p$ spin-up electrons, as the $3d$ spin-up bands are fully occupied. 

Another prominent feature of the \textit{ab-initio} spectra is that the timing of the resonance emission is shifted from the expected long/short trajectories. From Fig. \ref{fig:TDFT_gabor}(c), the resonant emission time in the spin-up channel follows roughly the typical long/short HHG semi-classical trajectories (as in the 1D model). However, in the more dominant spin-down channel the resonance emission is delayed by$\sim 0.18T\approx 480$ attoseconds. This result contradicts that from our 1D SAE model, suggesting an electronic interactions and spin-selective mechanism by which AIS resonances enhance HHG. Interestingly, the resonant emission yield builds up as the simulation progresses, while this build-up is absent in the non-resonant harmonics and in the spin-up channel, suggesting population of the AIS is playing a key role. A possible mechanism for this is given by the 4-step model in which a $3d$ electron first tunnels out (Cr$^+$ $3p^63d^5(^6S) \rightarrow$ Cr$^{2+} 3p^63d^4(^5D)+\epsilon l$) , gains kinetic energy in the laser field, and re-collides with Cr$^{+2}$ to excite a partial $3p$ spin-down electronic wavepacket into the Cr$^+ 3p^5(^2P)3d^6(^5D)(^6P)$ AIS, forming a superposition of AIS and ground state. This superposition has a time dependent dipole moment that oscillates at the frequency of H29. 
\begin{figure}[htb]
  \centering
  \includegraphics[width=0.9\textwidth]{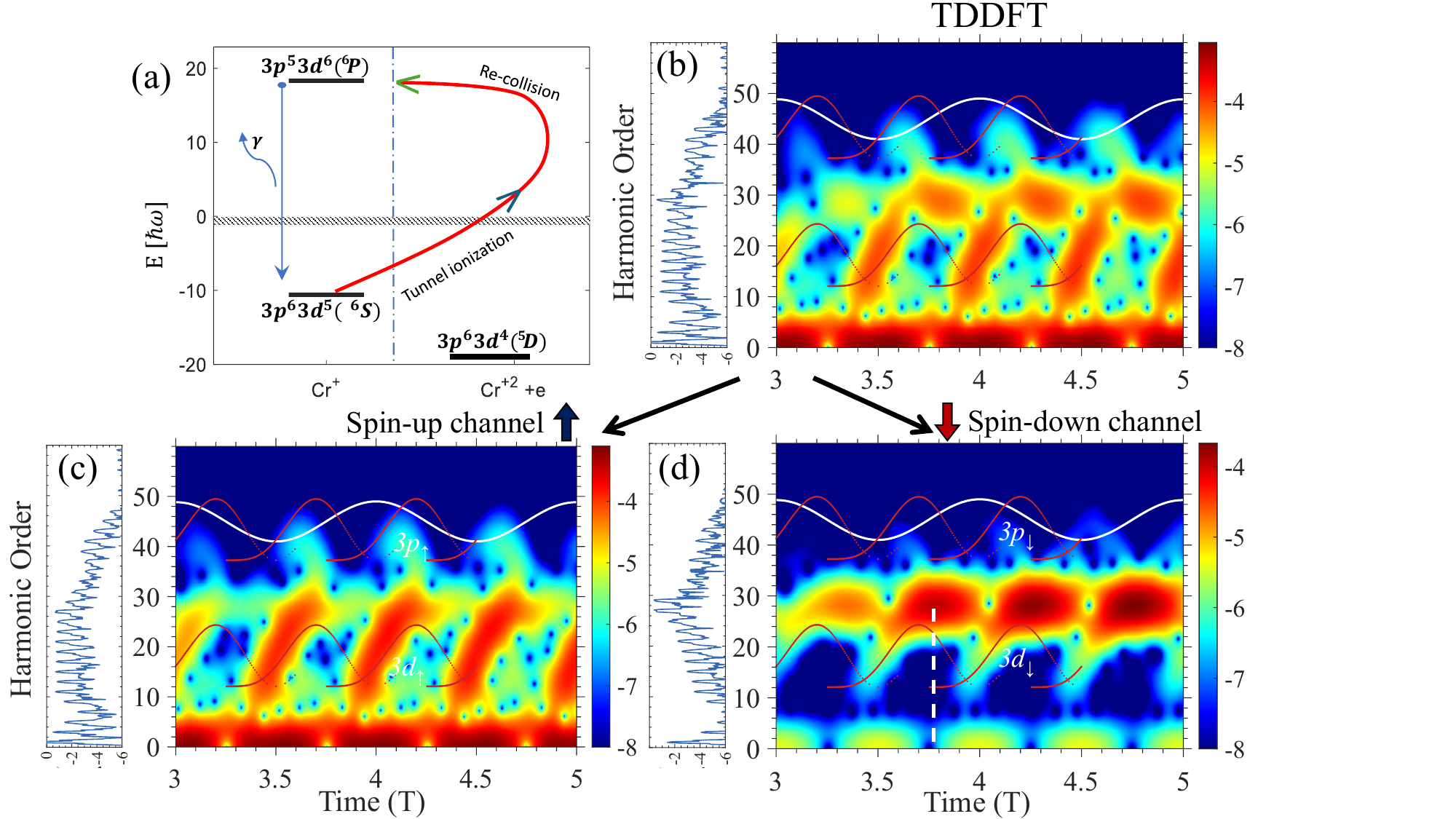}
  \caption{\textbf{rHHG with TDDFT.} \textbf{(a)} Energy level configuration in Cr$^{+}$. \textbf{(b)} (Right) Gabor transform of total calculated HHG emission. (Left) Corresponding HHG spectrum. Top and bottom red curves denote the emission time and energy according to the classical trajectories of the 3-step model (low energy curves associated to $3d$ electrons and high energy curves are associated to  $3p$ electrons). White lines illustrate the driving field. \textbf{(c)} Same as (a) but for spin-up response, roughly following the semiclassical model for $3d$ electrons. \textbf{(d)} Same as (c) but for spin-down response, where $3p$ spin-down states dominate rHHG (see color scales in (c) and (d)), which is delayed by 250 as relative to spin-up response.} \label{fig:TDFT_gabor}
\end{figure}

\begin{figure}[htb]
  \centering
  \includegraphics[width=0.9\textwidth, trim={1cm 0cm 2.2cm 1cm}, clip]{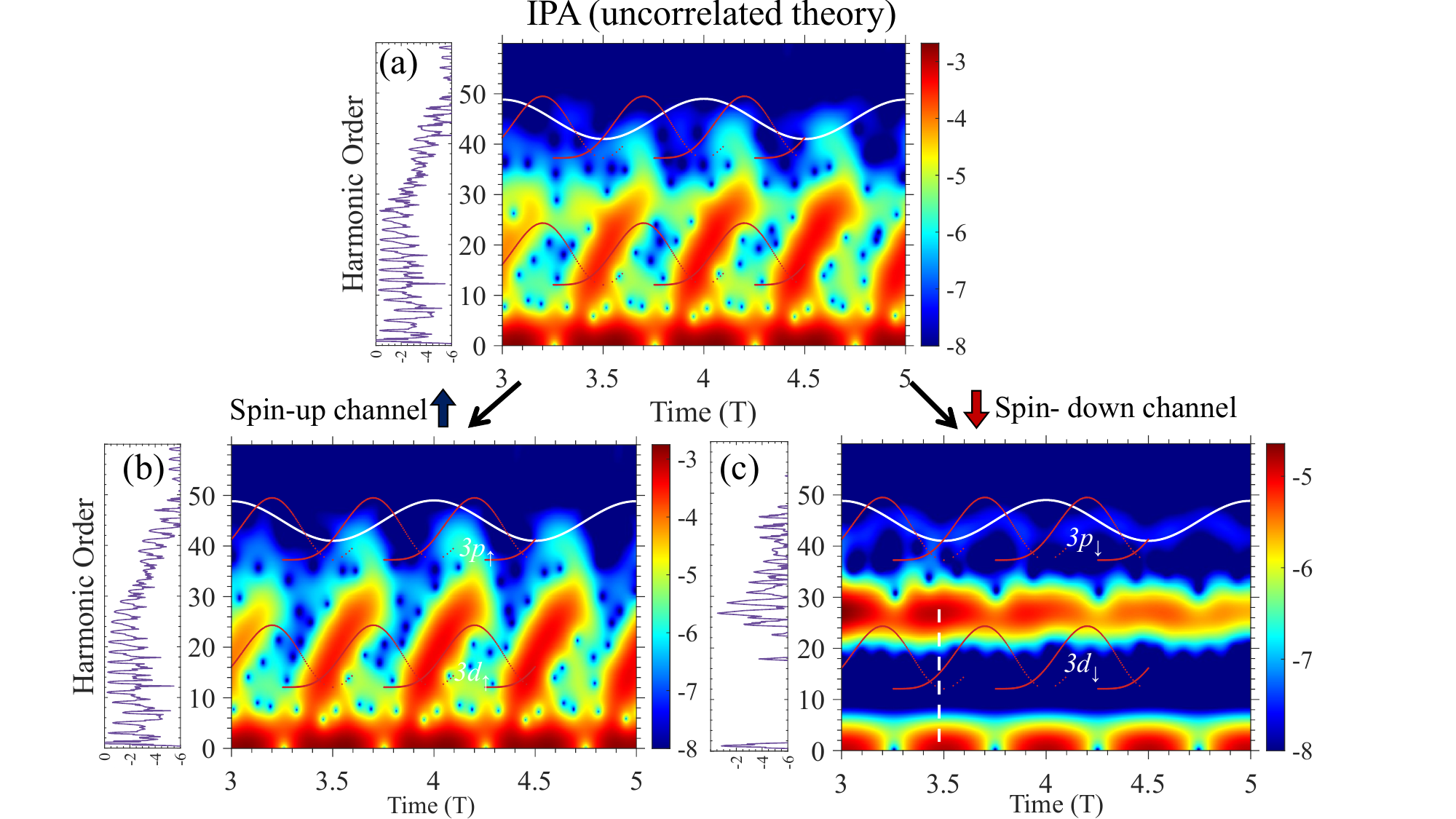}
  \caption{\textbf{rHHG within the IPA.} \textbf{(a-c)} Same as (b-d) in Fig.\ref{fig:TDFT_gabor}. Note the spin-down response on resonance is in-phase with the maxima/minima of the driving field and strongly suppressed, unlike in Fig.\ref{fig:TDFT_gabor}.} \label{fig:IPA_gabor}
\end{figure}

To further explore this conjecture, we invoke the IPA, i.e. freezing electronic interactions altogether. The results are shown in fig. \ref{fig:IPA_gabor} for both spin channels and the total emission, and reflect a prominent effect - upon freezing correlations the resonance enhancement is completely suppressed. Similar results were recently shown in another system\cite{romanov_simulation_2024}. Here however, the spin-up channel HHG response remains largely similar with/without the IPA, while the spin-down HHG response is substantially suppressed and is the source of the disappearance of rHHG. Notably, the emission from the spin-down electrons no longer follows the semi-classical trajectories of the $3d$ electrons, proving that the spin-down $3p\rightarrow3d$ transition is facilitated by dynamical electron correlations even in the tunneling regime. Further, while the resonance emission line still exists in the IPA spin-down channel, it is orders of magnitude weaker, has phase shifted, and is now aligned with the maximum of the electric field, similar to the perturbative low-order harmonics. In that respect, even though the atomic system still supports a resonant state in the absence of correlations, it does not contribute to rHHG since emission from typical HHG channels is more prominent, and the resonance emission mechanism differs from that of the correlated system. 

The spin-down emission spectra in both the IPA and correlated models closely resemble the 1D model, exhibiting a dip at lower harmonics followed by a peak at resonance. In the 1D case, this has been attributed to the returning electron’s reduced ability to penetrate the potential barrier \cite{tudorovskaya_high-order_2011}. 

Overall, we conclude that in the IPA, the $3p$ down states do not transition into $3d$ states, and the resonance is emitted by pure multi-photon transitions rather than excitation through the continuum.
In contrast, spin-down $3p$ electrons require electronic correlations in order to resonantly emit HHG in Cr$^{+}$ in experimental settings, following their transition into $3d$ levels, also incurring a time delay. These results are further corroborated by \textit{ab-initio} simulations for the single-photon absorption and ionization cross sections (see SI), supporting our conclusions that the down spin channel is the main contributor to absorption at the resonance.

\begin{figure}[h]
 \includegraphics[width=0.9\textwidth, trim={0 10.5cm 16.5cm 0cm}, clip]{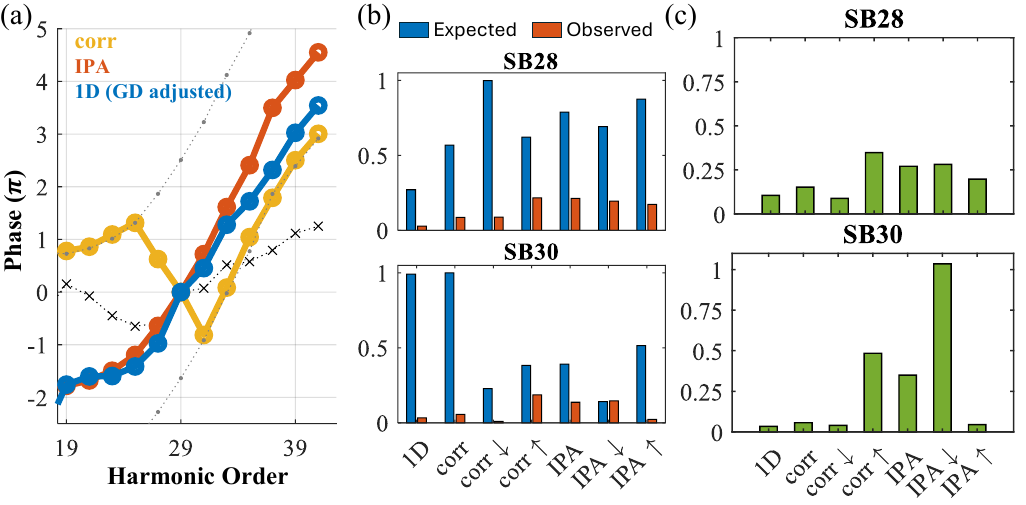}
\caption{\textbf{rHHG phase and RABBIT analysis.} \textbf{(a)} Harmonic phase comparison with varying levels of theory. Yellow and orange curves denote the fully correlated and IPA simulations, respectively. The blue line denotes the group-delay-corrected 1D simulation. Gray dotted lines with dot markers indicate the vertically shifted IPA, while the gray dotted line with $\times$ markers corresponds to the unadjusted 1D result. \textbf{(b)} Observed and expected SB visibility (for SB28 and SB30) for various levels of theory and spin channels. \textbf{(c)} Normalized visibility for both sidebands and various levels of theory. The correlated theory shows strongly reduced normalized visibility, corresponding to the strong phase jump in (a).} \label{fig:RABBIT_v2}
\end{figure}

Finally, we explore the phase of the harmonics at all levels of theory with a RABBIT-like simulation. We extract the harmonic phase across orders, as well as the RABBIT sideband (SB) visibility per order (see SI for details). The results presented in Fig. \ref{fig:RABBIT_v2}(a) show a sharp negative phase jump across the resonance that arises only in the correlated theory. This shift is up to $2.3\pi$ through the resonance, accompanied by a reduction in SB visibility. A vertically shifted IPA curve which aligns with the start and end of the correlated curve shows excellent agreement between the models, except near the resonance.
We have found that differences in the 1D and IPA simulations can be corrected for with group velocity delay corrections (blue line in Fig. \ref{fig:RABBIT_v2}(a), see SI for details), rendering both theories analogous. The phase jump in rHHG can potentially explain previous measurements of reduced SB visibility \cite{haessler_phase_nodate}, since it suggests a sharp phase jump even within the resonant harmonic itself (the $2.3\pi$ shift occurs abruptly over 3 harmonic orders).

To further test this, Fig. \ref{fig:RABBIT_v2}(b) presents the actual numerically-observed RABBIT SB visibility and the expected visibility due to two-source interference with unequal intensities. The normalized visibility is shown in Fig. \ref{fig:RABBIT_v2}(c). For both SB28 and SB30 (adjacent to the resonance), across all models, the expected visibility is larger than the observed with the exception of SB30 IPA spin-down, where they are roughly equal. For both SBs, the normalized correlated visibility is drastically lower than the normalized IPA visibility (SB28 - 15.1\% vs 27.1\%;  SB30 5.6\% vs 35\%). The spin-resolved components show similar behavior (most noticeably the spin-down component in SB30 (4\% vs 103.5\%)). An outlier is the spin-up component of SB30 (48.5\% vs 4.5\%), which we attribute to the extremely low harmonic signal in this channel. Interestingly, the 1D SAE model yields a normalized visibility similar to that of the spin-down component in the correlated model. This contrasts with the harmonic phase, which more closely resembles that of the IPA model. While this may seem unexpected, it is important to note that the 1D and IPA models are not equivalent. The 1D model employs a synthetic potential supporting a sharp shape resonance, which may differ from the IPA simulations, where this resonance could be less pronounced. Crucially, only the fully correlated theory accurately reproduces the correct harmonic phase near resonance,
also strongly reducing SB visibility in correspondence with measurements\cite{haessler_phase_nodate}.

In summary, we theoretically explored resonance-enhanced high harmonic generation (rHHG) in Cr$^{+}$ ions by employing three complementary approaches: (i) a 1D SAE model where the resonance is modeled by a shape-resonance, (ii) \textit{ab-initio} spin-time-dependent density functional theory (spin-TDDFT), and (iii) spin-TDDFT within the independent particle approximation (IPA) with electronic interactions artificially removed. 

The 1D model successfully reproduces experimental rHHG spectra, where we showed that the mechanism for rHHG in this case is tunnel-ionization followed by re-collision (without associated time delays). TDDFT also reproduces experimental spectra, but offers a more comprehensive view into rHHG physics --- we uncovered that the primary contribution to enhancement originates from spin-down $3p$ states rather than spin-up $3d$ electrons (that intuitively should dominate due to their lower ionization potential). Strikingly, the more dominant spin-down emission is delayed by $\sim$480 attoseconds compared to the spin-up $3d$ emission. These effects contrast with the 1D model and IPA simulations and point to a mechanism involving electron interaction and spin selectivity, also aligning with the 4-step picture: a $3d$ spin-up electron tunnel ionizes and accelerates in the continuum, re-collides with the parent ion, exciting a $3p$ spin-down electron (only possible through correlations), exciting an autoionization state. This proceeds to coherently decay to the ground state, emitting a resonantly-enhanced photon. Lastly, we explored RABBIT simulations that investigate the harmonic phases and sideband visibility. These simulations showed a sharp phase jump ($\sim2.3\pi$) across the resonance, only in the correlated theory. Correspondingly, normalized sideband visibility near the resonance was drastically lower in the correlated theory compared to the IPA (especially for the dominant spin-down component), which might explain earlier measurements\cite{haessler_phase_nodate}. 

Overall, our work establishes that the resonance enhancement mechanism is physically different in the absence of correlations, and thus suggests that SAE and semi-classical models must be phase and time-delay corrected due to electronic interactions. Our work resolves several open questions in the field, and should pave way to novel ultrafast spectroscopies of electron correlations and resonances.

\textit{Acknowledgments.} ON gratefully acknowledges the scientific support of Prof. Dr. Angel Rubio and the Young Faculty Award from the National Quantum Science and Technology program of Israel's Council of Higher Education Planning and Budgeting Committee. GM acknowledges support from the Israel Science Foundation Grant No. 65618 and Grant No. 89223.
AP acknowledges from the Israel Science Foundation  Grant No. 1484/24 and the Alon Fellowship of the Israeli Council of Higher Education.

\bibliography{references}

\begin{thebibliography}{8}%
\makeatletter
\providecommand \@ifxundefined [1]{%
 \@ifx{#1\undefined}
}%
\providecommand \@ifnum [1]{%
 \ifnum #1\expandafter \@firstoftwo
 \else \expandafter \@secondoftwo
 \fi
}%
\providecommand \@ifx [1]{%
 \ifx #1\expandafter \@firstoftwo
 \else \expandafter \@secondoftwo
 \fi
}%
\providecommand \natexlab [1]{#1}%
\providecommand \enquote  [1]{``#1''}%
\providecommand \bibnamefont  [1]{#1}%
\providecommand \bibfnamefont [1]{#1}%
\providecommand \citenamefont [1]{#1}%
\providecommand \href@noop [0]{\@secondoftwo}%
\providecommand \href [0]{\begingroup \@sanitize@url \@href}%
\providecommand \@href[1]{\@@startlink{#1}\@@href}%
\providecommand \@@href[1]{\endgroup#1\@@endlink}%
\providecommand \@sanitize@url [0]{\catcode `\\12\catcode `\$12\catcode `\&12\catcode `\#12\catcode `\^12\catcode `\_12\catcode `\%12\relax}%
\providecommand \@@startlink[1]{}%
\providecommand \@@endlink[0]{}%
\providecommand \url  [0]{\begingroup\@sanitize@url \@url }%
\providecommand \@url [1]{\endgroup\@href {#1}{\urlprefix }}%
\providecommand \urlprefix  [0]{URL }%
\providecommand \Eprint [0]{\href }%
\providecommand \doibase [0]{https://doi.org/}%
\providecommand \selectlanguage [0]{\@gobble}%
\providecommand \bibinfo  [0]{\@secondoftwo}%
\providecommand \bibfield  [0]{\@secondoftwo}%
\providecommand \translation [1]{[#1]}%
\providecommand \BibitemOpen [0]{}%
\providecommand \bibitemStop [0]{}%
\providecommand \bibitemNoStop [0]{.\EOS\space}%
\providecommand \EOS [0]{\spacefactor3000\relax}%
\providecommand \BibitemShut  [1]{\csname bibitem#1\endcsname}%
\let\auto@bib@innerbib\@empty
\bibitem [{MAT(2023)}]{MATLAB2023}%
  \BibitemOpen
  \href@noop {} {\bibinfo {title} {{The MathWorks Inc.}, matlab version: 23.2.0.2485118 (r2023b) update 6}},\ \bibinfo {howpublished} {\url{https://www.mathworks.com}} (\bibinfo {year} {2023})\BibitemShut {NoStop}%
\bibitem [{\citenamefont {Johnson}(2021)}]{johnson2021notesperfectlymatchedlayers}%
  \BibitemOpen
  \bibfield  {author} {\bibinfo {author} {\bibfnamefont {S.~G.}\ \bibnamefont {Johnson}},\ }\href {https://arxiv.org/abs/2108.05348} {\bibinfo {title} {Notes on perfectly matched layers (pmls)}} (\bibinfo {year} {2021}),\ \Eprint {https://arxiv.org/abs/2108.05348} {arXiv:2108.05348 [cs.CE]} \BibitemShut {NoStop}%
\bibitem [{\citenamefont {Moiseyev}(1998)}]{moiseyev_quantum_1998}%
  \BibitemOpen
  \bibfield  {author} {\bibinfo {author} {\bibfnamefont {N.}~\bibnamefont {Moiseyev}},\ }\bibfield  {title} {\bibinfo {title} {Quantum theory of resonances: calculating energies, widths and cross-sections by complex scaling},\ }\href {https://doi.org/10.1016/S0370-1573(98)00002-7} {\bibfield  {journal} {\bibinfo  {journal} {Physics Reports}\ }\textbf {\bibinfo {volume} {302}},\ \bibinfo {pages} {212} (\bibinfo {year} {1998})}\BibitemShut {NoStop}%
\bibitem [{\citenamefont {Aguilar}\ and\ \citenamefont {Combes}(1971)}]{aguilar_class_1971}%
  \BibitemOpen
  \bibfield  {author} {\bibinfo {author} {\bibfnamefont {J.}~\bibnamefont {Aguilar}}\ and\ \bibinfo {author} {\bibfnamefont {J.~M.}\ \bibnamefont {Combes}},\ }\bibfield  {title} {\bibinfo {title} {A class of analytic perturbations for one-body {Schrödinger} {Hamiltonians}},\ }\href {https://doi.org/10.1007/BF01877510} {\bibfield  {journal} {\bibinfo  {journal} {Communications in Mathematical Physics}\ }\textbf {\bibinfo {volume} {22}},\ \bibinfo {pages} {269} (\bibinfo {year} {1971})}\BibitemShut {NoStop}%
\bibitem [{\citenamefont {Miller}(2016)}]{miller2016}%
  \BibitemOpen
  \bibfield  {author} {\bibinfo {author} {\bibfnamefont {M.~R.}\ \bibnamefont {Miller}},\ }\href@noop {} {} (\bibinfo {year} {2016}),\ \bibinfo {note} {\emph{Time Resolving Electron Dynamics in Atomic and Molecular Systems Using High-Harmonic Spectroscopy}, {Ph.D.} thesis, University of Colorado Boulder}\BibitemShut {NoStop}%
\bibitem [{\citenamefont {Weiner}(2011)}]{weiner2011ultrafast}%
  \BibitemOpen
  \bibfield  {author} {\bibinfo {author} {\bibfnamefont {A.~M.}\ \bibnamefont {Weiner}},\ }\href@noop {} {\emph {\bibinfo {title} {Ultrafast optics}}}\ (\bibinfo  {publisher} {John Wiley \& Sons},\ \bibinfo {year} {2011})\BibitemShut {NoStop}%
\bibitem [{\citenamefont {Zavin}\ and\ \citenamefont {Moiseyev}(2004)}]{zavin_one-dimensional_2004}%
  \BibitemOpen
  \bibfield  {author} {\bibinfo {author} {\bibfnamefont {R.}~\bibnamefont {Zavin}}\ and\ \bibinfo {author} {\bibfnamefont {N.}~\bibnamefont {Moiseyev}},\ }\bibfield  {title} {\bibinfo {title} {One-dimensional symmetric rectangular well: from bound to resonance via self-orthogonal virtual state},\ }\href {https://doi.org/10.1088/0305-4470/37/16/011} {\bibfield  {journal} {\bibinfo  {journal} {Journal of Physics A: Mathematical and General}\ }\textbf {\bibinfo {volume} {37}},\ \bibinfo {pages} {4619} (\bibinfo {year} {2004})}\BibitemShut {NoStop}%
\bibitem [{\citenamefont {Nussenzveig}(1959)}]{nussenzveig_poles_1959}%
  \BibitemOpen
  \bibfield  {author} {\bibinfo {author} {\bibfnamefont {H.~M.}\ \bibnamefont {Nussenzveig}},\ }\bibfield  {title} {\bibinfo {title} {The poles of the {S}-matrix of a rectangular potential well of barrier},\ }\href {https://doi.org/10.1016/0029-5582(59)90293-7} {\bibfield  {journal} {\bibinfo  {journal} {Nuclear Physics}\ }\textbf {\bibinfo {volume} {11}},\ \bibinfo {pages} {499} (\bibinfo {year} {1959})}\BibitemShut {NoStop}%
\end{thebibliography}%


\begin{thebibliography}{63}%
\makeatletter
\providecommand \@ifxundefined [1]{%
 \@ifx{#1\undefined}
}%
\providecommand \@ifnum [1]{%
 \ifnum #1\expandafter \@firstoftwo
 \else \expandafter \@secondoftwo
 \fi
}%
\providecommand \@ifx [1]{%
 \ifx #1\expandafter \@firstoftwo
 \else \expandafter \@secondoftwo
 \fi
}%
\providecommand \natexlab [1]{#1}%
\providecommand \enquote  [1]{``#1''}%
\providecommand \bibnamefont  [1]{#1}%
\providecommand \bibfnamefont [1]{#1}%
\providecommand \citenamefont [1]{#1}%
\providecommand \href@noop [0]{\@secondoftwo}%
\providecommand \href [0]{\begingroup \@sanitize@url \@href}%
\providecommand \@href[1]{\@@startlink{#1}\@@href}%
\providecommand \@@href[1]{\endgroup#1\@@endlink}%
\providecommand \@sanitize@url [0]{\catcode `\\12\catcode `\$12\catcode `\&12\catcode `\#12\catcode `\^12\catcode `\_12\catcode `\%12\relax}%
\providecommand \@@startlink[1]{}%
\providecommand \@@endlink[0]{}%
\providecommand \url  [0]{\begingroup\@sanitize@url \@url }%
\providecommand \@url [1]{\endgroup\@href {#1}{\urlprefix }}%
\providecommand \urlprefix  [0]{URL }%
\providecommand \Eprint [0]{\href }%
\providecommand \doibase [0]{https://doi.org/}%
\providecommand \selectlanguage [0]{\@gobble}%
\providecommand \bibinfo  [0]{\@secondoftwo}%
\providecommand \bibfield  [0]{\@secondoftwo}%
\providecommand \translation [1]{[#1]}%
\providecommand \BibitemOpen [0]{}%
\providecommand \bibitemStop [0]{}%
\providecommand \bibitemNoStop [0]{.\EOS\space}%
\providecommand \EOS [0]{\spacefactor3000\relax}%
\providecommand \BibitemShut  [1]{\csname bibitem#1\endcsname}%
\let\auto@bib@innerbib\@empty
\bibitem [{\citenamefont {Hädrich}\ \emph {et~al.}(2015)\citenamefont {Hädrich}, \citenamefont {Krebs}, \citenamefont {Hoffmann}, \citenamefont {Klenke}, \citenamefont {Rothhardt}, \citenamefont {Limpert},\ and\ \citenamefont {Tünnermann}}]{hadrich_exploring_2015}%
  \BibitemOpen
  \bibfield  {author} {\bibinfo {author} {\bibfnamefont {S.}~\bibnamefont {Hädrich}}, \bibinfo {author} {\bibfnamefont {M.}~\bibnamefont {Krebs}}, \bibinfo {author} {\bibfnamefont {A.}~\bibnamefont {Hoffmann}}, \bibinfo {author} {\bibfnamefont {A.}~\bibnamefont {Klenke}}, \bibinfo {author} {\bibfnamefont {J.}~\bibnamefont {Rothhardt}}, \bibinfo {author} {\bibfnamefont {J.}~\bibnamefont {Limpert}},\ and\ \bibinfo {author} {\bibfnamefont {A.}~\bibnamefont {Tünnermann}},\ }\bibfield  {title} {\bibinfo {title} {Exploring new avenues in high repetition rate table-top coherent extreme ultraviolet sources},\ }\href {https://doi.org/10.1038/lsa.2015.93} {\bibfield  {journal} {\bibinfo  {journal} {Light: Science \& Applications}\ }\textbf {\bibinfo {volume} {4}},\ \bibinfo {pages} {e320} (\bibinfo {year} {2015})},\ \bibinfo {note} {publisher: Nature Publishing Group}\BibitemShut {NoStop}%
\bibitem [{\citenamefont {Chang}\ \emph {et~al.}(1997)\citenamefont {Chang}, \citenamefont {Rundquist}, \citenamefont {Wang}, \citenamefont {Murnane},\ and\ \citenamefont {Kapteyn}}]{chang_generation_1997}%
  \BibitemOpen
  \bibfield  {author} {\bibinfo {author} {\bibfnamefont {Z.}~\bibnamefont {Chang}}, \bibinfo {author} {\bibfnamefont {A.}~\bibnamefont {Rundquist}}, \bibinfo {author} {\bibfnamefont {H.}~\bibnamefont {Wang}}, \bibinfo {author} {\bibfnamefont {M.~M.}\ \bibnamefont {Murnane}},\ and\ \bibinfo {author} {\bibfnamefont {H.~C.}\ \bibnamefont {Kapteyn}},\ }\bibfield  {title} {\bibinfo {title} {Generation of {Coherent} {Soft} {X} {Rays} at 2.7 nm {Using} {High} {Harmonics}},\ }\href {https://doi.org/10.1103/PhysRevLett.79.2967} {\bibfield  {journal} {\bibinfo  {journal} {Physical Review Letters}\ }\textbf {\bibinfo {volume} {79}},\ \bibinfo {pages} {2967} (\bibinfo {year} {1997})},\ \bibinfo {note} {publisher: American Physical Society}\BibitemShut {NoStop}%
\bibitem [{\citenamefont {Spielmann}\ \emph {et~al.}(1997)\citenamefont {Spielmann}, \citenamefont {Burnett}, \citenamefont {Sartania}, \citenamefont {Koppitsch}, \citenamefont {Schnürer}, \citenamefont {Kan}, \citenamefont {Lenzner}, \citenamefont {Wobrauschek},\ and\ \citenamefont {Krausz}}]{spielmann_generation_1997}%
  \BibitemOpen
  \bibfield  {author} {\bibinfo {author} {\bibfnamefont {C.}~\bibnamefont {Spielmann}}, \bibinfo {author} {\bibfnamefont {N.~H.}\ \bibnamefont {Burnett}}, \bibinfo {author} {\bibfnamefont {S.}~\bibnamefont {Sartania}}, \bibinfo {author} {\bibfnamefont {R.}~\bibnamefont {Koppitsch}}, \bibinfo {author} {\bibfnamefont {M.}~\bibnamefont {Schnürer}}, \bibinfo {author} {\bibfnamefont {C.}~\bibnamefont {Kan}}, \bibinfo {author} {\bibfnamefont {M.}~\bibnamefont {Lenzner}}, \bibinfo {author} {\bibfnamefont {P.}~\bibnamefont {Wobrauschek}},\ and\ \bibinfo {author} {\bibfnamefont {F.}~\bibnamefont {Krausz}},\ }\bibfield  {title} {\bibinfo {title} {Generation of {Coherent} {X}-rays in the {Water} {Window} {Using} 5-{Femtosecond} {Laser} {Pulses}},\ }\href {https://doi.org/10.1126/science.278.5338.661} {\bibfield  {journal} {\bibinfo  {journal} {Science}\ }\textbf {\bibinfo {volume} {278}},\ \bibinfo {pages} {661} (\bibinfo {year} {1997})},\ \bibinfo {note} {publisher: American Association for the Advancement of
  Science}\BibitemShut {NoStop}%
\bibitem [{\citenamefont {Li}\ \emph {et~al.}(2020)\citenamefont {Li}, \citenamefont {Lu}, \citenamefont {Chew}, \citenamefont {Han}, \citenamefont {Li}, \citenamefont {Wu}, \citenamefont {Wang}, \citenamefont {Ghimire},\ and\ \citenamefont {Chang}}]{li_attosecond_2020}%
  \BibitemOpen
  \bibfield  {author} {\bibinfo {author} {\bibfnamefont {J.}~\bibnamefont {Li}}, \bibinfo {author} {\bibfnamefont {J.}~\bibnamefont {Lu}}, \bibinfo {author} {\bibfnamefont {A.}~\bibnamefont {Chew}}, \bibinfo {author} {\bibfnamefont {S.}~\bibnamefont {Han}}, \bibinfo {author} {\bibfnamefont {J.}~\bibnamefont {Li}}, \bibinfo {author} {\bibfnamefont {Y.}~\bibnamefont {Wu}}, \bibinfo {author} {\bibfnamefont {H.}~\bibnamefont {Wang}}, \bibinfo {author} {\bibfnamefont {S.}~\bibnamefont {Ghimire}},\ and\ \bibinfo {author} {\bibfnamefont {Z.}~\bibnamefont {Chang}},\ }\bibfield  {title} {\bibinfo {title} {Attosecond science based on high harmonic generation from gases and solids},\ }\href {https://doi.org/10.1038/s41467-020-16480-6} {\bibfield  {journal} {\bibinfo  {journal} {Nature Communications}\ }\textbf {\bibinfo {volume} {11}},\ \bibinfo {pages} {2748} (\bibinfo {year} {2020})},\ \bibinfo {note} {publisher: Nature Publishing Group}\BibitemShut {NoStop}%
\bibitem [{\citenamefont {Marcus}\ \emph {et~al.}(2012)\citenamefont {Marcus}, \citenamefont {Helml}, \citenamefont {Gu}, \citenamefont {Deng}, \citenamefont {Hartmann}, \citenamefont {Kobayashi}, \citenamefont {Strueder}, \citenamefont {Kienberger},\ and\ \citenamefont {Krausz}}]{marcus_subfemtosecond_2012}%
  \BibitemOpen
  \bibfield  {author} {\bibinfo {author} {\bibfnamefont {G.}~\bibnamefont {Marcus}}, \bibinfo {author} {\bibfnamefont {W.}~\bibnamefont {Helml}}, \bibinfo {author} {\bibfnamefont {X.}~\bibnamefont {Gu}}, \bibinfo {author} {\bibfnamefont {Y.}~\bibnamefont {Deng}}, \bibinfo {author} {\bibfnamefont {R.}~\bibnamefont {Hartmann}}, \bibinfo {author} {\bibfnamefont {T.}~\bibnamefont {Kobayashi}}, \bibinfo {author} {\bibfnamefont {L.}~\bibnamefont {Strueder}}, \bibinfo {author} {\bibfnamefont {R.}~\bibnamefont {Kienberger}},\ and\ \bibinfo {author} {\bibfnamefont {F.}~\bibnamefont {Krausz}},\ }\bibfield  {title} {\bibinfo {title} {Subfemtosecond k-{Shell} {Excitation} with a {Few}-{Cycle} {Infrared} {Laser} {Field}},\ }\href {https://doi.org/10.1103/PhysRevLett.108.023201} {\bibfield  {journal} {\bibinfo  {journal} {Physical Review Letters}\ }\textbf {\bibinfo {volume} {108}},\ \bibinfo {pages} {023201} (\bibinfo {year} {2012})},\ \bibinfo {note} {publisher: American Physical Society}\BibitemShut {NoStop}%
\bibitem [{\citenamefont {Shafir}\ \emph {et~al.}(2012)\citenamefont {Shafir}, \citenamefont {Soifer}, \citenamefont {Bruner}, \citenamefont {Dagan}, \citenamefont {Mairesse}, \citenamefont {Patchkovskii}, \citenamefont {Ivanov}, \citenamefont {Smirnova},\ and\ \citenamefont {Dudovich}}]{shafir_resolving_2012}%
  \BibitemOpen
  \bibfield  {author} {\bibinfo {author} {\bibfnamefont {D.}~\bibnamefont {Shafir}}, \bibinfo {author} {\bibfnamefont {H.}~\bibnamefont {Soifer}}, \bibinfo {author} {\bibfnamefont {B.~D.}\ \bibnamefont {Bruner}}, \bibinfo {author} {\bibfnamefont {M.}~\bibnamefont {Dagan}}, \bibinfo {author} {\bibfnamefont {Y.}~\bibnamefont {Mairesse}}, \bibinfo {author} {\bibfnamefont {S.}~\bibnamefont {Patchkovskii}}, \bibinfo {author} {\bibfnamefont {M.~Y.}\ \bibnamefont {Ivanov}}, \bibinfo {author} {\bibfnamefont {O.}~\bibnamefont {Smirnova}},\ and\ \bibinfo {author} {\bibfnamefont {N.}~\bibnamefont {Dudovich}},\ }\bibfield  {title} {\bibinfo {title} {Resolving the time when an electron exits a tunnelling barrier},\ }\href {https://doi.org/10.1038/nature11025} {\bibfield  {journal} {\bibinfo  {journal} {Nature}\ }\textbf {\bibinfo {volume} {485}},\ \bibinfo {pages} {343} (\bibinfo {year} {2012})},\ \bibinfo {note} {publisher: Nature Publishing Group}\BibitemShut {NoStop}%
\bibitem [{\citenamefont {Smirnova}\ \emph {et~al.}(2009)\citenamefont {Smirnova}, \citenamefont {Mairesse}, \citenamefont {Patchkovskii}, \citenamefont {Dudovich}, \citenamefont {Villeneuve}, \citenamefont {Corkum},\ and\ \citenamefont {Ivanov}}]{smirnova_high_2009}%
  \BibitemOpen
  \bibfield  {author} {\bibinfo {author} {\bibfnamefont {O.}~\bibnamefont {Smirnova}}, \bibinfo {author} {\bibfnamefont {Y.}~\bibnamefont {Mairesse}}, \bibinfo {author} {\bibfnamefont {S.}~\bibnamefont {Patchkovskii}}, \bibinfo {author} {\bibfnamefont {N.}~\bibnamefont {Dudovich}}, \bibinfo {author} {\bibfnamefont {D.}~\bibnamefont {Villeneuve}}, \bibinfo {author} {\bibfnamefont {P.}~\bibnamefont {Corkum}},\ and\ \bibinfo {author} {\bibfnamefont {M.~Y.}\ \bibnamefont {Ivanov}},\ }\bibfield  {title} {\bibinfo {title} {High harmonic interferometry of multi-electron dynamics in molecules},\ }\href {https://doi.org/10.1038/nature08253} {\bibfield  {journal} {\bibinfo  {journal} {Nature}\ }\textbf {\bibinfo {volume} {460}},\ \bibinfo {pages} {972} (\bibinfo {year} {2009})},\ \bibinfo {note} {publisher: Nature Publishing Group}\BibitemShut {NoStop}%
\bibitem [{\citenamefont {Veisz}\ \emph {et~al.}(2013)\citenamefont {Veisz}, \citenamefont {Rivas}, \citenamefont {Marcus}, \citenamefont {Gu}, \citenamefont {Cardenas}, \citenamefont {Mikhailova}, \citenamefont {Buck}, \citenamefont {Wittmann}, \citenamefont {Sears}, \citenamefont {Chou},\ and\ \citenamefont {{others}}}]{veisz_generation_2013}%
  \BibitemOpen
  \bibfield  {author} {\bibinfo {author} {\bibfnamefont {L.}~\bibnamefont {Veisz}}, \bibinfo {author} {\bibfnamefont {D.}~\bibnamefont {Rivas}}, \bibinfo {author} {\bibfnamefont {G.}~\bibnamefont {Marcus}}, \bibinfo {author} {\bibfnamefont {X.}~\bibnamefont {Gu}}, \bibinfo {author} {\bibfnamefont {D.}~\bibnamefont {Cardenas}}, \bibinfo {author} {\bibfnamefont {J.}~\bibnamefont {Mikhailova}}, \bibinfo {author} {\bibfnamefont {A.}~\bibnamefont {Buck}}, \bibinfo {author} {\bibfnamefont {T.}~\bibnamefont {Wittmann}}, \bibinfo {author} {\bibfnamefont {C.~M.~S.}\ \bibnamefont {Sears}}, \bibinfo {author} {\bibfnamefont {S.-W.}\ \bibnamefont {Chou}},\ and\ \bibinfo {author} {\bibnamefont {{others}}},\ }\bibfield  {title} {\bibinfo {title} {Generation and applications of sub-5-fs multi-10-{TW} light pulses},\ }in\ \href {http://www.osapublishing.org/abstract.cfm?uri=CLEOPR-2013-TuD2_3} {\emph {\bibinfo {booktitle} {Conference on {Lasers} and {Electro}-{Optics}/{Pacific} {Rim}}}}\ (\bibinfo  {publisher} {Optical
  Society of America},\ \bibinfo {year} {2013})\ p.\ \bibinfo {pages} {TuD2\_3},\ \bibinfo {note} {00005}\BibitemShut {NoStop}%
\bibitem [{\citenamefont {Bergues}\ \emph {et~al.}(2018)\citenamefont {Bergues}, \citenamefont {Rivas}, \citenamefont {Weidman}, \citenamefont {Muschet}, \citenamefont {Helml}, \citenamefont {Guggenmos}, \citenamefont {Pervak}, \citenamefont {Kleineberg}, \citenamefont {Marcus}, \citenamefont {Kienberger}, \citenamefont {Charalambidis}, \citenamefont {Tzallas}, \citenamefont {Schröder}, \citenamefont {Krausz},\ and\ \citenamefont {Veisz}}]{bergues_tabletop_2018}%
  \BibitemOpen
  \bibfield  {author} {\bibinfo {author} {\bibfnamefont {B.}~\bibnamefont {Bergues}}, \bibinfo {author} {\bibfnamefont {D.~E.}\ \bibnamefont {Rivas}}, \bibinfo {author} {\bibfnamefont {M.}~\bibnamefont {Weidman}}, \bibinfo {author} {\bibfnamefont {A.~A.}\ \bibnamefont {Muschet}}, \bibinfo {author} {\bibfnamefont {W.}~\bibnamefont {Helml}}, \bibinfo {author} {\bibfnamefont {A.}~\bibnamefont {Guggenmos}}, \bibinfo {author} {\bibfnamefont {V.}~\bibnamefont {Pervak}}, \bibinfo {author} {\bibfnamefont {U.}~\bibnamefont {Kleineberg}}, \bibinfo {author} {\bibfnamefont {G.}~\bibnamefont {Marcus}}, \bibinfo {author} {\bibfnamefont {R.}~\bibnamefont {Kienberger}}, \bibinfo {author} {\bibfnamefont {D.}~\bibnamefont {Charalambidis}}, \bibinfo {author} {\bibfnamefont {P.}~\bibnamefont {Tzallas}}, \bibinfo {author} {\bibfnamefont {H.}~\bibnamefont {Schröder}}, \bibinfo {author} {\bibfnamefont {F.}~\bibnamefont {Krausz}},\ and\ \bibinfo {author} {\bibfnamefont {L.}~\bibnamefont {Veisz}},\ }\bibfield  {title} {\bibinfo
  {title} {Tabletop nonlinear optics in the 100-{eV} spectral region},\ }\href {https://doi.org/10.1364/OPTICA.5.000237} {\bibfield  {journal} {\bibinfo  {journal} {Optica}\ }\textbf {\bibinfo {volume} {5}},\ \bibinfo {pages} {237} (\bibinfo {year} {2018})}\BibitemShut {NoStop}%
\bibitem [{\citenamefont {Sobolev}\ \emph {et~al.}(2024)\citenamefont {Sobolev}, \citenamefont {Volkov}, \citenamefont {Svirplys}, \citenamefont {Thomas}, \citenamefont {Witting}, \citenamefont {Vrakking},\ and\ \citenamefont {Schütte}}]{sobolev_terawatt-level_2024}%
  \BibitemOpen
  \bibfield  {author} {\bibinfo {author} {\bibfnamefont {E.}~\bibnamefont {Sobolev}}, \bibinfo {author} {\bibfnamefont {M.}~\bibnamefont {Volkov}}, \bibinfo {author} {\bibfnamefont {E.}~\bibnamefont {Svirplys}}, \bibinfo {author} {\bibfnamefont {J.}~\bibnamefont {Thomas}}, \bibinfo {author} {\bibfnamefont {T.}~\bibnamefont {Witting}}, \bibinfo {author} {\bibfnamefont {M.~J.~J.}\ \bibnamefont {Vrakking}},\ and\ \bibinfo {author} {\bibfnamefont {B.}~\bibnamefont {Schütte}},\ }\bibfield  {title} {\bibinfo {title} {Terawatt-level three-stage pulse compression for all-attosecond pump-probe spectroscopy},\ }\href {https://doi.org/10.1364/OE.540265} {\bibfield  {journal} {\bibinfo  {journal} {Optics Express}\ }\textbf {\bibinfo {volume} {32}},\ \bibinfo {pages} {46251} (\bibinfo {year} {2024})},\ \bibinfo {note} {publisher: Optica Publishing Group}\BibitemShut {NoStop}%
\bibitem [{\citenamefont {Mondal}\ \emph {et~al.}(2023)\citenamefont {Mondal}, \citenamefont {Neufeld}, \citenamefont {Yin}, \citenamefont {Nourbakhsh}, \citenamefont {Svoboda}, \citenamefont {Rubio}, \citenamefont {Tancogne-Dejean},\ and\ \citenamefont {Wörner}}]{mondal_high-harmonic_2023}%
  \BibitemOpen
  \bibfield  {author} {\bibinfo {author} {\bibfnamefont {A.}~\bibnamefont {Mondal}}, \bibinfo {author} {\bibfnamefont {O.}~\bibnamefont {Neufeld}}, \bibinfo {author} {\bibfnamefont {Z.}~\bibnamefont {Yin}}, \bibinfo {author} {\bibfnamefont {Z.}~\bibnamefont {Nourbakhsh}}, \bibinfo {author} {\bibfnamefont {V.}~\bibnamefont {Svoboda}}, \bibinfo {author} {\bibfnamefont {A.}~\bibnamefont {Rubio}}, \bibinfo {author} {\bibfnamefont {N.}~\bibnamefont {Tancogne-Dejean}},\ and\ \bibinfo {author} {\bibfnamefont {H.~J.}\ \bibnamefont {Wörner}},\ }\bibfield  {title} {\bibinfo {title} {High-harmonic spectroscopy of low-energy electron-scattering dynamics in liquids},\ }\href {https://doi.org/10.1038/s41567-023-02214-0} {\bibfield  {journal} {\bibinfo  {journal} {Nature Physics}\ }\textbf {\bibinfo {volume} {19}},\ \bibinfo {pages} {1813} (\bibinfo {year} {2023})},\ \bibinfo {note} {publisher: Nature Publishing Group}\BibitemShut {NoStop}%
\bibitem [{\citenamefont {Deng}\ \emph {et~al.}(2020)\citenamefont {Deng}, \citenamefont {Zeng}, \citenamefont {Komm}, \citenamefont {Zheng}, \citenamefont {Helml}, \citenamefont {Xie}, \citenamefont {Filus}, \citenamefont {Dumergue}, \citenamefont {Flender}, \citenamefont {Kurucz}, \citenamefont {Haizer}, \citenamefont {Kiss}, \citenamefont {Kahaly}, \citenamefont {Li},\ and\ \citenamefont {Marcus}}]{deng_laser-induced_2020}%
  \BibitemOpen
  \bibfield  {author} {\bibinfo {author} {\bibfnamefont {Y.}~\bibnamefont {Deng}}, \bibinfo {author} {\bibfnamefont {Z.}~\bibnamefont {Zeng}}, \bibinfo {author} {\bibfnamefont {P.}~\bibnamefont {Komm}}, \bibinfo {author} {\bibfnamefont {Y.}~\bibnamefont {Zheng}}, \bibinfo {author} {\bibfnamefont {W.}~\bibnamefont {Helml}}, \bibinfo {author} {\bibfnamefont {X.}~\bibnamefont {Xie}}, \bibinfo {author} {\bibfnamefont {Z.}~\bibnamefont {Filus}}, \bibinfo {author} {\bibfnamefont {M.}~\bibnamefont {Dumergue}}, \bibinfo {author} {\bibfnamefont {R.}~\bibnamefont {Flender}}, \bibinfo {author} {\bibfnamefont {M.}~\bibnamefont {Kurucz}}, \bibinfo {author} {\bibfnamefont {L.}~\bibnamefont {Haizer}}, \bibinfo {author} {\bibfnamefont {B.}~\bibnamefont {Kiss}}, \bibinfo {author} {\bibfnamefont {S.}~\bibnamefont {Kahaly}}, \bibinfo {author} {\bibfnamefont {R.}~\bibnamefont {Li}},\ and\ \bibinfo {author} {\bibfnamefont {G.}~\bibnamefont {Marcus}},\ }\bibfield  {title} {\bibinfo {title} {Laser-induced inner-shell excitations
  through direct electron re-collision versus indirect collision},\ }\href {https://doi.org/10.1364/OE.395927} {\bibfield  {journal} {\bibinfo  {journal} {Optics Express}\ }\textbf {\bibinfo {volume} {28}},\ \bibinfo {pages} {23251} (\bibinfo {year} {2020})},\ \bibinfo {note} {publisher: Optical Society of America}\BibitemShut {NoStop}%
\bibitem [{\citenamefont {Alcalà}\ \emph {et~al.}(2022)\citenamefont {Alcalà}, \citenamefont {Bhattacharya}, \citenamefont {Biegert}, \citenamefont {Ciappina}, \citenamefont {Elu}, \citenamefont {Graß}, \citenamefont {Grochowski}, \citenamefont {Lewenstein}, \citenamefont {Palau}, \citenamefont {Sidiropoulos}, \citenamefont {Steinle},\ and\ \citenamefont {Tyulnev}}]{alcala_high-harmonic_2022}%
  \BibitemOpen
  \bibfield  {author} {\bibinfo {author} {\bibfnamefont {J.}~\bibnamefont {Alcalà}}, \bibinfo {author} {\bibfnamefont {U.}~\bibnamefont {Bhattacharya}}, \bibinfo {author} {\bibfnamefont {J.}~\bibnamefont {Biegert}}, \bibinfo {author} {\bibfnamefont {M.}~\bibnamefont {Ciappina}}, \bibinfo {author} {\bibfnamefont {U.}~\bibnamefont {Elu}}, \bibinfo {author} {\bibfnamefont {T.}~\bibnamefont {Graß}}, \bibinfo {author} {\bibfnamefont {P.~T.}\ \bibnamefont {Grochowski}}, \bibinfo {author} {\bibfnamefont {M.}~\bibnamefont {Lewenstein}}, \bibinfo {author} {\bibfnamefont {A.}~\bibnamefont {Palau}}, \bibinfo {author} {\bibfnamefont {T.~P.~H.}\ \bibnamefont {Sidiropoulos}}, \bibinfo {author} {\bibfnamefont {T.}~\bibnamefont {Steinle}},\ and\ \bibinfo {author} {\bibfnamefont {I.}~\bibnamefont {Tyulnev}},\ }\bibfield  {title} {\bibinfo {title} {High-harmonic spectroscopy of quantum phase transitions in a high-{Tc} superconductor},\ }\href {https://doi.org/10.1073/pnas.2207766119} {\bibfield  {journal} {\bibinfo
  {journal} {Proceedings of the National Academy of Sciences}\ }\textbf {\bibinfo {volume} {119}},\ \bibinfo {pages} {e2207766119} (\bibinfo {year} {2022})},\ \bibinfo {note} {publisher: Proceedings of the National Academy of Sciences}\BibitemShut {NoStop}%
\bibitem [{\citenamefont {Neufeld}\ and\ \citenamefont {Cohen}(2020)}]{neufeld_probing_2020}%
  \BibitemOpen
  \bibfield  {author} {\bibinfo {author} {\bibfnamefont {O.}~\bibnamefont {Neufeld}}\ and\ \bibinfo {author} {\bibfnamefont {O.}~\bibnamefont {Cohen}},\ }\bibfield  {title} {\bibinfo {title} {Probing ultrafast electron correlations in high harmonic generation},\ }\href {https://doi.org/10.1103/PhysRevResearch.2.033037} {\bibfield  {journal} {\bibinfo  {journal} {Physical Review Research}\ }\textbf {\bibinfo {volume} {2}},\ \bibinfo {pages} {033037} (\bibinfo {year} {2020})},\ \bibinfo {note} {publisher: American Physical Society}\BibitemShut {NoStop}%
\bibitem [{\citenamefont {Neufeld}\ \emph {et~al.}(2019{\natexlab{a}})\citenamefont {Neufeld}, \citenamefont {Ayuso}, \citenamefont {Decleva}, \citenamefont {Ivanov}, \citenamefont {Smirnova},\ and\ \citenamefont {Cohen}}]{neufeld_ultrasensitive_2019}%
  \BibitemOpen
  \bibfield  {author} {\bibinfo {author} {\bibfnamefont {O.}~\bibnamefont {Neufeld}}, \bibinfo {author} {\bibfnamefont {D.}~\bibnamefont {Ayuso}}, \bibinfo {author} {\bibfnamefont {P.}~\bibnamefont {Decleva}}, \bibinfo {author} {\bibfnamefont {M.~Y.}\ \bibnamefont {Ivanov}}, \bibinfo {author} {\bibfnamefont {O.}~\bibnamefont {Smirnova}},\ and\ \bibinfo {author} {\bibfnamefont {O.}~\bibnamefont {Cohen}},\ }\bibfield  {title} {\bibinfo {title} {Ultrasensitive {Chiral} {Spectroscopy} by {Dynamical} {Symmetry} {Breaking} in {High} {Harmonic} {Generation}},\ }\href {https://doi.org/10.1103/PhysRevX.9.031002} {\bibfield  {journal} {\bibinfo  {journal} {Physical Review X}\ }\textbf {\bibinfo {volume} {9}},\ \bibinfo {pages} {031002} (\bibinfo {year} {2019}{\natexlab{a}})},\ \bibinfo {note} {publisher: American Physical Society}\BibitemShut {NoStop}%
\bibitem [{\citenamefont {Hamer}\ \emph {et~al.}(2022)\citenamefont {Hamer}, \citenamefont {Mauger}, \citenamefont {Folorunso}, \citenamefont {Lopata}, \citenamefont {Jones}, \citenamefont {DiMauro}, \citenamefont {Schafer},\ and\ \citenamefont {Gaarde}}]{hamer_characterizing_2022}%
  \BibitemOpen
  \bibfield  {author} {\bibinfo {author} {\bibfnamefont {K.~A.}\ \bibnamefont {Hamer}}, \bibinfo {author} {\bibfnamefont {F.}~\bibnamefont {Mauger}}, \bibinfo {author} {\bibfnamefont {A.~S.}\ \bibnamefont {Folorunso}}, \bibinfo {author} {\bibfnamefont {K.}~\bibnamefont {Lopata}}, \bibinfo {author} {\bibfnamefont {R.~R.}\ \bibnamefont {Jones}}, \bibinfo {author} {\bibfnamefont {L.~F.}\ \bibnamefont {DiMauro}}, \bibinfo {author} {\bibfnamefont {K.~J.}\ \bibnamefont {Schafer}},\ and\ \bibinfo {author} {\bibfnamefont {M.~B.}\ \bibnamefont {Gaarde}},\ }\bibfield  {title} {\bibinfo {title} {Characterizing particle-like charge-migration dynamics with high-order harmonic sideband spectroscopy},\ }\href {https://doi.org/10.1103/PhysRevA.106.013103} {\bibfield  {journal} {\bibinfo  {journal} {Physical Review A}\ }\textbf {\bibinfo {volume} {106}},\ \bibinfo {pages} {013103} (\bibinfo {year} {2022})},\ \bibinfo {note} {publisher: American Physical Society}\BibitemShut {NoStop}%
\bibitem [{\citenamefont {Freudenstein}\ \emph {et~al.}(2022)\citenamefont {Freudenstein}, \citenamefont {Borsch}, \citenamefont {Meierhofer}, \citenamefont {Afanasiev}, \citenamefont {Schmid}, \citenamefont {Sandner}, \citenamefont {Liebich}, \citenamefont {Girnghuber}, \citenamefont {Knorr}, \citenamefont {Kira},\ and\ \citenamefont {Huber}}]{freudenstein_attosecond_2022}%
  \BibitemOpen
  \bibfield  {author} {\bibinfo {author} {\bibfnamefont {J.}~\bibnamefont {Freudenstein}}, \bibinfo {author} {\bibfnamefont {M.}~\bibnamefont {Borsch}}, \bibinfo {author} {\bibfnamefont {M.}~\bibnamefont {Meierhofer}}, \bibinfo {author} {\bibfnamefont {D.}~\bibnamefont {Afanasiev}}, \bibinfo {author} {\bibfnamefont {C.~P.}\ \bibnamefont {Schmid}}, \bibinfo {author} {\bibfnamefont {F.}~\bibnamefont {Sandner}}, \bibinfo {author} {\bibfnamefont {M.}~\bibnamefont {Liebich}}, \bibinfo {author} {\bibfnamefont {A.}~\bibnamefont {Girnghuber}}, \bibinfo {author} {\bibfnamefont {M.}~\bibnamefont {Knorr}}, \bibinfo {author} {\bibfnamefont {M.}~\bibnamefont {Kira}},\ and\ \bibinfo {author} {\bibfnamefont {R.}~\bibnamefont {Huber}},\ }\bibfield  {title} {\bibinfo {title} {Attosecond clocking of correlations between {Bloch} electrons},\ }\href {https://doi.org/10.1038/s41586-022-05190-2} {\bibfield  {journal} {\bibinfo  {journal} {Nature}\ }\textbf {\bibinfo {volume} {610}},\ \bibinfo {pages} {290} (\bibinfo {year}
  {2022})},\ \bibinfo {note} {publisher: Nature Publishing Group}\BibitemShut {NoStop}%
\bibitem [{\citenamefont {Neufeld}\ \emph {et~al.}(2022{\natexlab{a}})\citenamefont {Neufeld}, \citenamefont {Zhang}, \citenamefont {De~Giovannini}, \citenamefont {Hübener},\ and\ \citenamefont {Rubio}}]{neufeld_probing_2022}%
  \BibitemOpen
  \bibfield  {author} {\bibinfo {author} {\bibfnamefont {O.}~\bibnamefont {Neufeld}}, \bibinfo {author} {\bibfnamefont {J.}~\bibnamefont {Zhang}}, \bibinfo {author} {\bibfnamefont {U.}~\bibnamefont {De~Giovannini}}, \bibinfo {author} {\bibfnamefont {H.}~\bibnamefont {Hübener}},\ and\ \bibinfo {author} {\bibfnamefont {A.}~\bibnamefont {Rubio}},\ }\bibfield  {title} {\bibinfo {title} {Probing phonon dynamics with multidimensional high harmonic carrier-envelope-phase spectroscopy},\ }\href {https://doi.org/10.1073/pnas.2204219119} {\bibfield  {journal} {\bibinfo  {journal} {Proceedings of the National Academy of Sciences}\ }\textbf {\bibinfo {volume} {119}},\ \bibinfo {pages} {e2204219119} (\bibinfo {year} {2022}{\natexlab{a}})},\ \bibinfo {note} {publisher: Proceedings of the National Academy of Sciences}\BibitemShut {NoStop}%
\bibitem [{\citenamefont {Heide}\ \emph {et~al.}(2022)\citenamefont {Heide}, \citenamefont {Kobayashi}, \citenamefont {Baykusheva}, \citenamefont {Jain}, \citenamefont {Sobota}, \citenamefont {Hashimoto}, \citenamefont {Kirchmann}, \citenamefont {Oh}, \citenamefont {Heinz}, \citenamefont {Reis},\ and\ \citenamefont {Ghimire}}]{heide_probing_2022}%
  \BibitemOpen
  \bibfield  {author} {\bibinfo {author} {\bibfnamefont {C.}~\bibnamefont {Heide}}, \bibinfo {author} {\bibfnamefont {Y.}~\bibnamefont {Kobayashi}}, \bibinfo {author} {\bibfnamefont {D.~R.}\ \bibnamefont {Baykusheva}}, \bibinfo {author} {\bibfnamefont {D.}~\bibnamefont {Jain}}, \bibinfo {author} {\bibfnamefont {J.~A.}\ \bibnamefont {Sobota}}, \bibinfo {author} {\bibfnamefont {M.}~\bibnamefont {Hashimoto}}, \bibinfo {author} {\bibfnamefont {P.~S.}\ \bibnamefont {Kirchmann}}, \bibinfo {author} {\bibfnamefont {S.}~\bibnamefont {Oh}}, \bibinfo {author} {\bibfnamefont {T.~F.}\ \bibnamefont {Heinz}}, \bibinfo {author} {\bibfnamefont {D.~A.}\ \bibnamefont {Reis}},\ and\ \bibinfo {author} {\bibfnamefont {S.}~\bibnamefont {Ghimire}},\ }\bibfield  {title} {\bibinfo {title} {Probing topological phase transitions using high-harmonic generation},\ }\href {https://doi.org/10.1038/s41566-022-01050-7} {\bibfield  {journal} {\bibinfo  {journal} {Nature Photonics}\ }\textbf {\bibinfo {volume} {16}},\ \bibinfo {pages} {620}
  (\bibinfo {year} {2022})},\ \bibinfo {note} {publisher: Nature Publishing Group}\BibitemShut {NoStop}%
\bibitem [{\citenamefont {Schmid}\ \emph {et~al.}(2021)\citenamefont {Schmid}, \citenamefont {Weigl}, \citenamefont {Grössing}, \citenamefont {Junk}, \citenamefont {Gorini}, \citenamefont {Schlauderer}, \citenamefont {Ito}, \citenamefont {Meierhofer}, \citenamefont {Hofmann}, \citenamefont {Afanasiev}, \citenamefont {Crewse}, \citenamefont {Kokh}, \citenamefont {Tereshchenko}, \citenamefont {Güdde}, \citenamefont {Evers}, \citenamefont {Wilhelm}, \citenamefont {Richter}, \citenamefont {Höfer},\ and\ \citenamefont {Huber}}]{schmid_tunable_2021}%
  \BibitemOpen
  \bibfield  {author} {\bibinfo {author} {\bibfnamefont {C.~P.}\ \bibnamefont {Schmid}}, \bibinfo {author} {\bibfnamefont {L.}~\bibnamefont {Weigl}}, \bibinfo {author} {\bibfnamefont {P.}~\bibnamefont {Grössing}}, \bibinfo {author} {\bibfnamefont {V.}~\bibnamefont {Junk}}, \bibinfo {author} {\bibfnamefont {C.}~\bibnamefont {Gorini}}, \bibinfo {author} {\bibfnamefont {S.}~\bibnamefont {Schlauderer}}, \bibinfo {author} {\bibfnamefont {S.}~\bibnamefont {Ito}}, \bibinfo {author} {\bibfnamefont {M.}~\bibnamefont {Meierhofer}}, \bibinfo {author} {\bibfnamefont {N.}~\bibnamefont {Hofmann}}, \bibinfo {author} {\bibfnamefont {D.}~\bibnamefont {Afanasiev}}, \bibinfo {author} {\bibfnamefont {J.}~\bibnamefont {Crewse}}, \bibinfo {author} {\bibfnamefont {K.~A.}\ \bibnamefont {Kokh}}, \bibinfo {author} {\bibfnamefont {O.~E.}\ \bibnamefont {Tereshchenko}}, \bibinfo {author} {\bibfnamefont {J.}~\bibnamefont {Güdde}}, \bibinfo {author} {\bibfnamefont {F.}~\bibnamefont {Evers}}, \bibinfo {author} {\bibfnamefont
  {J.}~\bibnamefont {Wilhelm}}, \bibinfo {author} {\bibfnamefont {K.}~\bibnamefont {Richter}}, \bibinfo {author} {\bibfnamefont {U.}~\bibnamefont {Höfer}},\ and\ \bibinfo {author} {\bibfnamefont {R.}~\bibnamefont {Huber}},\ }\bibfield  {title} {\bibinfo {title} {Tunable non-integer high-harmonic generation in a topological insulator},\ }\href {https://doi.org/10.1038/s41586-021-03466-7} {\bibfield  {journal} {\bibinfo  {journal} {Nature}\ }\textbf {\bibinfo {volume} {593}},\ \bibinfo {pages} {385} (\bibinfo {year} {2021})},\ \bibinfo {note} {publisher: Nature Publishing Group}\BibitemShut {NoStop}%
\bibitem [{\citenamefont {Zhang}\ \emph {et~al.}(2024)\citenamefont {Zhang}, \citenamefont {Wang}, \citenamefont {Lengers}, \citenamefont {Wigger}, \citenamefont {Reiter}, \citenamefont {Kuhn}, \citenamefont {Wörner},\ and\ \citenamefont {Luu}}]{zhang_high-harmonic_2024}%
  \BibitemOpen
  \bibfield  {author} {\bibinfo {author} {\bibfnamefont {J.}~\bibnamefont {Zhang}}, \bibinfo {author} {\bibfnamefont {Z.}~\bibnamefont {Wang}}, \bibinfo {author} {\bibfnamefont {F.}~\bibnamefont {Lengers}}, \bibinfo {author} {\bibfnamefont {D.}~\bibnamefont {Wigger}}, \bibinfo {author} {\bibfnamefont {D.~E.}\ \bibnamefont {Reiter}}, \bibinfo {author} {\bibfnamefont {T.}~\bibnamefont {Kuhn}}, \bibinfo {author} {\bibfnamefont {H.~J.}\ \bibnamefont {Wörner}},\ and\ \bibinfo {author} {\bibfnamefont {T.~T.}\ \bibnamefont {Luu}},\ }\bibfield  {title} {\bibinfo {title} {High-harmonic spectroscopy probes lattice dynamics},\ }\href {https://doi.org/10.1038/s41566-024-01457-4} {\bibfield  {journal} {\bibinfo  {journal} {Nature Photonics}\ }\textbf {\bibinfo {volume} {18}},\ \bibinfo {pages} {792} (\bibinfo {year} {2024})},\ \bibinfo {note} {publisher: Nature Publishing Group}\BibitemShut {NoStop}%
\bibitem [{\citenamefont {Krause}\ \emph {et~al.}(1992)\citenamefont {Krause}, \citenamefont {Schafer},\ and\ \citenamefont {Kulander}}]{krause_high-order_1992}%
  \BibitemOpen
  \bibfield  {author} {\bibinfo {author} {\bibfnamefont {J.~L.}\ \bibnamefont {Krause}}, \bibinfo {author} {\bibfnamefont {K.~J.}\ \bibnamefont {Schafer}},\ and\ \bibinfo {author} {\bibfnamefont {K.~C.}\ \bibnamefont {Kulander}},\ }\bibfield  {title} {\bibinfo {title} {High-order harmonic generation from atoms and ions in the high intensity regime},\ }\href {https://doi.org/10.1103/PhysRevLett.68.3535} {\bibfield  {journal} {\bibinfo  {journal} {Physical Review Letters}\ }\textbf {\bibinfo {volume} {68}},\ \bibinfo {pages} {3535} (\bibinfo {year} {1992})}\BibitemShut {NoStop}%
\bibitem [{\citenamefont {Corkum}(1993)}]{corkum_plasma_1993}%
  \BibitemOpen
  \bibfield  {author} {\bibinfo {author} {\bibfnamefont {P.~B.}\ \bibnamefont {Corkum}},\ }\bibfield  {title} {\bibinfo {title} {Plasma perspective on strong field multiphoton ionization},\ }\href {https://doi.org/10.1103/PhysRevLett.71.1994} {\bibfield  {journal} {\bibinfo  {journal} {Physical Review Letters}\ }\textbf {\bibinfo {volume} {71}},\ \bibinfo {pages} {1994} (\bibinfo {year} {1993})}\BibitemShut {NoStop}%
\bibitem [{\citenamefont {He}\ \emph {et~al.}(2022)\citenamefont {He}, \citenamefont {Sun}, \citenamefont {Lan}, \citenamefont {He}, \citenamefont {Wang}, \citenamefont {Wang}, \citenamefont {Zhu}, \citenamefont {Li}, \citenamefont {Cao}, \citenamefont {Lu},\ and\ \citenamefont {Lin}}]{he_filming_2022}%
  \BibitemOpen
  \bibfield  {author} {\bibinfo {author} {\bibfnamefont {L.}~\bibnamefont {He}}, \bibinfo {author} {\bibfnamefont {S.}~\bibnamefont {Sun}}, \bibinfo {author} {\bibfnamefont {P.}~\bibnamefont {Lan}}, \bibinfo {author} {\bibfnamefont {Y.}~\bibnamefont {He}}, \bibinfo {author} {\bibfnamefont {B.}~\bibnamefont {Wang}}, \bibinfo {author} {\bibfnamefont {P.}~\bibnamefont {Wang}}, \bibinfo {author} {\bibfnamefont {X.}~\bibnamefont {Zhu}}, \bibinfo {author} {\bibfnamefont {L.}~\bibnamefont {Li}}, \bibinfo {author} {\bibfnamefont {W.}~\bibnamefont {Cao}}, \bibinfo {author} {\bibfnamefont {P.}~\bibnamefont {Lu}},\ and\ \bibinfo {author} {\bibfnamefont {C.~D.}\ \bibnamefont {Lin}},\ }\bibfield  {title} {\bibinfo {title} {Filming movies of attosecond charge migration in single molecules with high harmonic spectroscopy},\ }\href {https://doi.org/10.1038/s41467-022-32313-0} {\bibfield  {journal} {\bibinfo  {journal} {Nature Communications}\ }\textbf {\bibinfo {volume} {13}},\ \bibinfo {pages} {4595} (\bibinfo {year}
  {2022})},\ \bibinfo {note} {publisher: Nature Publishing Group}\BibitemShut {NoStop}%
\bibitem [{\citenamefont {Baykusheva}\ \emph {et~al.}(2016)\citenamefont {Baykusheva}, \citenamefont {Ahsan}, \citenamefont {Lin},\ and\ \citenamefont {Wörner}}]{baykusheva_bicircular_2016}%
  \BibitemOpen
  \bibfield  {author} {\bibinfo {author} {\bibfnamefont {D.}~\bibnamefont {Baykusheva}}, \bibinfo {author} {\bibfnamefont {M.~S.}\ \bibnamefont {Ahsan}}, \bibinfo {author} {\bibfnamefont {N.}~\bibnamefont {Lin}},\ and\ \bibinfo {author} {\bibfnamefont {H.~J.}\ \bibnamefont {Wörner}},\ }\bibfield  {title} {\bibinfo {title} {Bicircular {High}-{Harmonic} {Spectroscopy} {Reveals} {Dynamical} {Symmetries} of {Atoms} and {Molecules}},\ }\href {https://doi.org/10.1103/PhysRevLett.116.123001} {\bibfield  {journal} {\bibinfo  {journal} {Physical Review Letters}\ }\textbf {\bibinfo {volume} {116}},\ \bibinfo {pages} {123001} (\bibinfo {year} {2016})},\ \bibinfo {note} {publisher: American Physical Society}\BibitemShut {NoStop}%
\bibitem [{\citenamefont {Frumker}\ \emph {et~al.}(2012)\citenamefont {Frumker}, \citenamefont {Hebeisen}, \citenamefont {Kajumba}, \citenamefont {Bertrand}, \citenamefont {Wörner}, \citenamefont {Spanner}, \citenamefont {Villeneuve}, \citenamefont {Naumov},\ and\ \citenamefont {Corkum}}]{frumker_oriented_2012}%
  \BibitemOpen
  \bibfield  {author} {\bibinfo {author} {\bibfnamefont {E.}~\bibnamefont {Frumker}}, \bibinfo {author} {\bibfnamefont {C.~T.}\ \bibnamefont {Hebeisen}}, \bibinfo {author} {\bibfnamefont {N.}~\bibnamefont {Kajumba}}, \bibinfo {author} {\bibfnamefont {J.~B.}\ \bibnamefont {Bertrand}}, \bibinfo {author} {\bibfnamefont {H.~J.}\ \bibnamefont {Wörner}}, \bibinfo {author} {\bibfnamefont {M.}~\bibnamefont {Spanner}}, \bibinfo {author} {\bibfnamefont {D.~M.}\ \bibnamefont {Villeneuve}}, \bibinfo {author} {\bibfnamefont {A.}~\bibnamefont {Naumov}},\ and\ \bibinfo {author} {\bibfnamefont {P.~B.}\ \bibnamefont {Corkum}},\ }\bibfield  {title} {\bibinfo {title} {Oriented {Rotational} {Wave}-{Packet} {Dynamics} {Studies} via {High} {Harmonic} {Generation}},\ }\href {https://doi.org/10.1103/PhysRevLett.109.113901} {\bibfield  {journal} {\bibinfo  {journal} {Physical Review Letters}\ }\textbf {\bibinfo {volume} {109}},\ \bibinfo {pages} {113901} (\bibinfo {year} {2012})},\ \bibinfo {note} {publisher: American Physical
  Society}\BibitemShut {NoStop}%
\bibitem [{\citenamefont {Neufeld}\ \emph {et~al.}(2022{\natexlab{b}})\citenamefont {Neufeld}, \citenamefont {Wengrowicz}, \citenamefont {Peleg}, \citenamefont {Rubio},\ and\ \citenamefont {Cohen}}]{neufeld_detecting_2022}%
  \BibitemOpen
  \bibfield  {author} {\bibinfo {author} {\bibfnamefont {O.}~\bibnamefont {Neufeld}}, \bibinfo {author} {\bibfnamefont {O.}~\bibnamefont {Wengrowicz}}, \bibinfo {author} {\bibfnamefont {O.}~\bibnamefont {Peleg}}, \bibinfo {author} {\bibfnamefont {A.}~\bibnamefont {Rubio}},\ and\ \bibinfo {author} {\bibfnamefont {O.}~\bibnamefont {Cohen}},\ }\bibfield  {title} {\bibinfo {title} {Detecting multiple chiral centers in chiral molecules with high harmonic generation},\ }\href {https://doi.org/10.1364/OE.445743} {\bibfield  {journal} {\bibinfo  {journal} {Optics Express}\ }\textbf {\bibinfo {volume} {30}},\ \bibinfo {pages} {3729} (\bibinfo {year} {2022}{\natexlab{b}})},\ \bibinfo {note} {publisher: Optica Publishing Group}\BibitemShut {NoStop}%
\bibitem [{\citenamefont {Saito}\ \emph {et~al.}(2017)\citenamefont {Saito}, \citenamefont {Xia}, \citenamefont {Lu}, \citenamefont {Kanai}, \citenamefont {Itatani},\ and\ \citenamefont {Ishii}}]{saito_observation_2017}%
  \BibitemOpen
  \bibfield  {author} {\bibinfo {author} {\bibfnamefont {N.}~\bibnamefont {Saito}}, \bibinfo {author} {\bibfnamefont {P.}~\bibnamefont {Xia}}, \bibinfo {author} {\bibfnamefont {F.}~\bibnamefont {Lu}}, \bibinfo {author} {\bibfnamefont {T.}~\bibnamefont {Kanai}}, \bibinfo {author} {\bibfnamefont {J.}~\bibnamefont {Itatani}},\ and\ \bibinfo {author} {\bibfnamefont {N.}~\bibnamefont {Ishii}},\ }\bibfield  {title} {\bibinfo {title} {Observation of selection rules for circularly polarized fields in high-harmonic generation from a crystalline solid},\ }\href {https://doi.org/10.1364/OPTICA.4.001333} {\bibfield  {journal} {\bibinfo  {journal} {Optica}\ }\textbf {\bibinfo {volume} {4}},\ \bibinfo {pages} {1333} (\bibinfo {year} {2017})},\ \bibinfo {note} {publisher: Optica Publishing Group}\BibitemShut {NoStop}%
\bibitem [{\citenamefont {Neufeld}\ \emph {et~al.}(2019{\natexlab{b}})\citenamefont {Neufeld}, \citenamefont {Podolsky},\ and\ \citenamefont {Cohen}}]{neufeld_floquet_2019}%
  \BibitemOpen
  \bibfield  {author} {\bibinfo {author} {\bibfnamefont {O.}~\bibnamefont {Neufeld}}, \bibinfo {author} {\bibfnamefont {D.}~\bibnamefont {Podolsky}},\ and\ \bibinfo {author} {\bibfnamefont {O.}~\bibnamefont {Cohen}},\ }\bibfield  {title} {\bibinfo {title} {Floquet group theory and its application to selection rules in harmonic generation},\ }\href {https://doi.org/10.1038/s41467-018-07935-y} {\bibfield  {journal} {\bibinfo  {journal} {Nature Communications}\ }\textbf {\bibinfo {volume} {10}},\ \bibinfo {pages} {405} (\bibinfo {year} {2019}{\natexlab{b}})},\ \bibinfo {note} {publisher: Nature Publishing Group}\BibitemShut {NoStop}%
\bibitem [{\citenamefont {Yuan}\ and\ \citenamefont {Bandrauk}(2018)}]{PhysRevA.97.023408}%
  \BibitemOpen
  \bibfield  {author} {\bibinfo {author} {\bibfnamefont {K.-J.}\ \bibnamefont {Yuan}}\ and\ \bibinfo {author} {\bibfnamefont {A.~D.}\ \bibnamefont {Bandrauk}},\ }\bibfield  {title} {\bibinfo {title} {Symmetry in circularly polarized molecular high-order harmonic generation with intense bicircular laser pulses},\ }\href {https://doi.org/10.1103/PhysRevA.97.023408} {\bibfield  {journal} {\bibinfo  {journal} {Phys. Rev. A}\ }\textbf {\bibinfo {volume} {97}},\ \bibinfo {pages} {023408} (\bibinfo {year} {2018})}\BibitemShut {NoStop}%
\bibitem [{\citenamefont {Nalda}\ \emph {et~al.}(2004)\citenamefont {Nalda}, \citenamefont {Heesel}, \citenamefont {Lein}, \citenamefont {Hay}, \citenamefont {Velotta}, \citenamefont {Springate}, \citenamefont {Castillejo},\ and\ \citenamefont {Marangos}}]{PhysRevA.69.031804}%
  \BibitemOpen
  \bibfield  {author} {\bibinfo {author} {\bibfnamefont {R.~d.}\ \bibnamefont {Nalda}}, \bibinfo {author} {\bibfnamefont {E.}~\bibnamefont {Heesel}}, \bibinfo {author} {\bibfnamefont {M.}~\bibnamefont {Lein}}, \bibinfo {author} {\bibfnamefont {N.}~\bibnamefont {Hay}}, \bibinfo {author} {\bibfnamefont {R.}~\bibnamefont {Velotta}}, \bibinfo {author} {\bibfnamefont {E.}~\bibnamefont {Springate}}, \bibinfo {author} {\bibfnamefont {M.}~\bibnamefont {Castillejo}},\ and\ \bibinfo {author} {\bibfnamefont {J.~P.}\ \bibnamefont {Marangos}},\ }\bibfield  {title} {\bibinfo {title} {Role of orbital symmetry in high-order harmonic generation from aligned molecules},\ }\href {https://doi.org/10.1103/PhysRevA.69.031804} {\bibfield  {journal} {\bibinfo  {journal} {Phys. Rev. A}\ }\textbf {\bibinfo {volume} {69}},\ \bibinfo {pages} {031804} (\bibinfo {year} {2004})}\BibitemShut {NoStop}%
\bibitem [{\citenamefont {Torres}\ \emph {et~al.}(2007)\citenamefont {Torres}, \citenamefont {Kajumba}, \citenamefont {Underwood}, \citenamefont {Robinson}, \citenamefont {Baker}, \citenamefont {Tisch}, \citenamefont {de~Nalda}, \citenamefont {Bryan}, \citenamefont {Velotta}, \citenamefont {Altucci}, \citenamefont {Turcu},\ and\ \citenamefont {Marangos}}]{torres_probing_2007}%
  \BibitemOpen
  \bibfield  {author} {\bibinfo {author} {\bibfnamefont {R.}~\bibnamefont {Torres}}, \bibinfo {author} {\bibfnamefont {N.}~\bibnamefont {Kajumba}}, \bibinfo {author} {\bibfnamefont {J.~G.}\ \bibnamefont {Underwood}}, \bibinfo {author} {\bibfnamefont {J.~S.}\ \bibnamefont {Robinson}}, \bibinfo {author} {\bibfnamefont {S.}~\bibnamefont {Baker}}, \bibinfo {author} {\bibfnamefont {J.~W.~G.}\ \bibnamefont {Tisch}}, \bibinfo {author} {\bibfnamefont {R.}~\bibnamefont {de~Nalda}}, \bibinfo {author} {\bibfnamefont {W.~A.}\ \bibnamefont {Bryan}}, \bibinfo {author} {\bibfnamefont {R.}~\bibnamefont {Velotta}}, \bibinfo {author} {\bibfnamefont {C.}~\bibnamefont {Altucci}}, \bibinfo {author} {\bibfnamefont {I.~C.~E.}\ \bibnamefont {Turcu}},\ and\ \bibinfo {author} {\bibfnamefont {J.~P.}\ \bibnamefont {Marangos}},\ }\bibfield  {title} {\bibinfo {title} {Probing {Orbital} {Structure} of {Polyatomic} {Molecules} by {High}-{Order} {Harmonic} {Generation}},\ }\href {https://doi.org/10.1103/PhysRevLett.98.203007} {\bibfield
  {journal} {\bibinfo  {journal} {Physical Review Letters}\ }\textbf {\bibinfo {volume} {98}},\ \bibinfo {pages} {203007} (\bibinfo {year} {2007})},\ \bibinfo {note} {publisher: American Physical Society}\BibitemShut {NoStop}%
\bibitem [{\citenamefont {Madsen}\ and\ \citenamefont {Madsen}(2007)}]{madsen_theoretical_2007}%
  \BibitemOpen
  \bibfield  {author} {\bibinfo {author} {\bibfnamefont {C.~B.}\ \bibnamefont {Madsen}}\ and\ \bibinfo {author} {\bibfnamefont {L.~B.}\ \bibnamefont {Madsen}},\ }\bibfield  {title} {\bibinfo {title} {Theoretical studies of high-order harmonic generation: {Effects} of symmetry, degeneracy, and orientation},\ }\href {https://doi.org/10.1103/PhysRevA.76.043419} {\bibfield  {journal} {\bibinfo  {journal} {Physical Review A}\ }\textbf {\bibinfo {volume} {76}},\ \bibinfo {pages} {043419} (\bibinfo {year} {2007})},\ \bibinfo {note} {publisher: American Physical Society}\BibitemShut {NoStop}%
\bibitem [{\citenamefont {Suzuki}\ \emph {et~al.}(2006)\citenamefont {Suzuki}, \citenamefont {Baba}, \citenamefont {Ganeev}, \citenamefont {Kuroda},\ and\ \citenamefont {Ozaki}}]{suzuki_anomalous_2006}%
  \BibitemOpen
  \bibfield  {author} {\bibinfo {author} {\bibfnamefont {M.}~\bibnamefont {Suzuki}}, \bibinfo {author} {\bibfnamefont {M.}~\bibnamefont {Baba}}, \bibinfo {author} {\bibfnamefont {R.}~\bibnamefont {Ganeev}}, \bibinfo {author} {\bibfnamefont {H.}~\bibnamefont {Kuroda}},\ and\ \bibinfo {author} {\bibfnamefont {T.}~\bibnamefont {Ozaki}},\ }\bibfield  {title} {\bibinfo {title} {Anomalous enhancement of a single high-order harmonic by using a laser-ablation tin plume at 47 nm},\ }\href {https://doi.org/10.1364/OL.31.003306} {\bibfield  {journal} {\bibinfo  {journal} {Optics Letters}\ }\textbf {\bibinfo {volume} {31}},\ \bibinfo {pages} {3306} (\bibinfo {year} {2006})}\BibitemShut {NoStop}%
\bibitem [{\citenamefont {Ganeev}(2009)}]{ganeev_generation_2009}%
  \BibitemOpen
  \bibfield  {author} {\bibinfo {author} {\bibfnamefont {R.~A.}\ \bibnamefont {Ganeev}},\ }\bibfield  {title} {\bibinfo {title} {Generation of high-order harmonics of high-power lasers in plasmas produced under irradiation of solid target surfaces by a prepulse},\ }\href {https://doi.org/10.3367/UFNe.0179.200901c.0065} {\bibfield  {journal} {\bibinfo  {journal} {Physics-Uspekhi}\ }\textbf {\bibinfo {volume} {52}},\ \bibinfo {pages} {55} (\bibinfo {year} {2009})}\BibitemShut {NoStop}%
\bibitem [{\citenamefont {Rosenthal}\ and\ \citenamefont {Marcus}(2015)}]{rosenthal_discriminating_2015}%
  \BibitemOpen
  \bibfield  {author} {\bibinfo {author} {\bibfnamefont {N.}~\bibnamefont {Rosenthal}}\ and\ \bibinfo {author} {\bibfnamefont {G.}~\bibnamefont {Marcus}},\ }\bibfield  {title} {\bibinfo {title} {Discriminating between the {Role} of {Phase} {Matching} and that of the {Single}-{Atom} {Response} in {Resonance} {Plasma}-{Plume} {High}-{Order} {Harmonic} {Generation}},\ }\href {https://doi.org/10.1103/PhysRevLett.115.133901} {\bibfield  {journal} {\bibinfo  {journal} {Physical Review Letters}\ }\textbf {\bibinfo {volume} {115}},\ \bibinfo {pages} {133901} (\bibinfo {year} {2015})}\BibitemShut {NoStop}%
\bibitem [{\citenamefont {Costello}\ \emph {et~al.}(1991)\citenamefont {Costello}, \citenamefont {Kennedy}, \citenamefont {Sonntag},\ and\ \citenamefont {Clark}}]{costello_3p_1991}%
  \BibitemOpen
  \bibfield  {author} {\bibinfo {author} {\bibfnamefont {J.~T.}\ \bibnamefont {Costello}}, \bibinfo {author} {\bibfnamefont {E.~T.}\ \bibnamefont {Kennedy}}, \bibinfo {author} {\bibfnamefont {B.~F.}\ \bibnamefont {Sonntag}},\ and\ \bibinfo {author} {\bibfnamefont {C.~W.}\ \bibnamefont {Clark}},\ }\bibfield  {title} {\bibinfo {title} {3p photoabsorption of free and bound {Cr}, {Cr}+, {Mn}, and {Mn}+},\ }\href {https://doi.org/10.1103/PhysRevA.43.1441} {\bibfield  {journal} {\bibinfo  {journal} {Physical Review A}\ }\textbf {\bibinfo {volume} {43}},\ \bibinfo {pages} {1441} (\bibinfo {year} {1991})}\BibitemShut {NoStop}%
\bibitem [{\citenamefont {Osawa}\ \emph {et~al.}(2012)\citenamefont {Osawa}, \citenamefont {Kawajiri}, \citenamefont {Suzuki}, \citenamefont {Nagata}, \citenamefont {Azuma},\ and\ \citenamefont {Koike}}]{osawa_photoion-yield_2012}%
  \BibitemOpen
  \bibfield  {author} {\bibinfo {author} {\bibfnamefont {T.}~\bibnamefont {Osawa}}, \bibinfo {author} {\bibfnamefont {K.}~\bibnamefont {Kawajiri}}, \bibinfo {author} {\bibfnamefont {N.}~\bibnamefont {Suzuki}}, \bibinfo {author} {\bibfnamefont {T.}~\bibnamefont {Nagata}}, \bibinfo {author} {\bibfnamefont {Y.}~\bibnamefont {Azuma}},\ and\ \bibinfo {author} {\bibfnamefont {F.}~\bibnamefont {Koike}},\ }\bibfield  {title} {\bibinfo {title} {Photoion-yield study of the 3p–3d giant resonance excitation region of isolated {Cr}, {Mn} and {Fe} atoms},\ }\href {https://doi.org/10.1088/0953-4075/45/22/225204} {\bibfield  {journal} {\bibinfo  {journal} {Journal of Physics B: Atomic, Molecular and Optical Physics}\ }\textbf {\bibinfo {volume} {45}},\ \bibinfo {pages} {225204} (\bibinfo {year} {2012})},\ \bibinfo {note} {00002}\BibitemShut {NoStop}%
\bibitem [{\citenamefont {Frolov}\ \emph {et~al.}(2010)\citenamefont {Frolov}, \citenamefont {Manakov},\ and\ \citenamefont {Starace}}]{frolov_potential_2010}%
  \BibitemOpen
  \bibfield  {author} {\bibinfo {author} {\bibfnamefont {M.~V.}\ \bibnamefont {Frolov}}, \bibinfo {author} {\bibfnamefont {N.~L.}\ \bibnamefont {Manakov}},\ and\ \bibinfo {author} {\bibfnamefont {A.~F.}\ \bibnamefont {Starace}},\ }\bibfield  {title} {\bibinfo {title} {Potential barrier effects in high-order harmonic generation by transition-metal ions},\ }\href {https://doi.org/10.1103/PhysRevA.82.023424} {\bibfield  {journal} {\bibinfo  {journal} {Physical Review A}\ }\textbf {\bibinfo {volume} {82}},\ \bibinfo {pages} {023424} (\bibinfo {year} {2010})},\ \bibinfo {note} {00037}\BibitemShut {NoStop}%
\bibitem [{\citenamefont {Kheifets}(2025)}]{kheifets_resonant_2025}%
  \BibitemOpen
  \bibfield  {author} {\bibinfo {author} {\bibfnamefont {A.~S.}\ \bibnamefont {Kheifets}},\ }\bibfield  {title} {\bibinfo {title} {Resonant photoionization and time delay},\ }\href {https://doi.org/10.1088/1361-6455/adbfea} {\bibfield  {journal} {\bibinfo  {journal} {Journal of Physics B: Atomic, Molecular and Optical Physics}\ }\textbf {\bibinfo {volume} {58}},\ \bibinfo {pages} {072001} (\bibinfo {year} {2025})}\BibitemShut {NoStop}%
\bibitem [{\citenamefont {Strelkov}(2010)}]{strelkov_role_2010}%
  \BibitemOpen
  \bibfield  {author} {\bibinfo {author} {\bibfnamefont {V.}~\bibnamefont {Strelkov}},\ }\bibfield  {title} {\bibinfo {title} {Role of {Autoionizing} {State} in {Resonant} {High}-{Order} {Harmonic} {Generation} and {Attosecond} {Pulse} {Production}},\ }\href {https://doi.org/10.1103/PhysRevLett.104.123901} {\bibfield  {journal} {\bibinfo  {journal} {Physical Review Letters}\ }\textbf {\bibinfo {volume} {104}},\ \bibinfo {pages} {123901} (\bibinfo {year} {2010})}\BibitemShut {NoStop}%
\bibitem [{\citenamefont {Strelkov}\ \emph {et~al.}(2014)\citenamefont {Strelkov}, \citenamefont {Khokhlova},\ and\ \citenamefont {Shubin}}]{strelkov_high-order_2014}%
  \BibitemOpen
  \bibfield  {author} {\bibinfo {author} {\bibfnamefont {V.~V.}\ \bibnamefont {Strelkov}}, \bibinfo {author} {\bibfnamefont {M.~A.}\ \bibnamefont {Khokhlova}},\ and\ \bibinfo {author} {\bibfnamefont {N.~Y.}\ \bibnamefont {Shubin}},\ }\bibfield  {title} {\bibinfo {title} {High-order harmonic generation and {Fano} resonances},\ }\href {https://doi.org/10.1103/PhysRevA.89.053833} {\bibfield  {journal} {\bibinfo  {journal} {Physical Review A}\ }\textbf {\bibinfo {volume} {89}},\ \bibinfo {pages} {053833} (\bibinfo {year} {2014})}\BibitemShut {NoStop}%
\bibitem [{\citenamefont {Tudorovskaya}\ and\ \citenamefont {Lein}(2011)}]{tudorovskaya_high-order_2011}%
  \BibitemOpen
  \bibfield  {author} {\bibinfo {author} {\bibfnamefont {M.}~\bibnamefont {Tudorovskaya}}\ and\ \bibinfo {author} {\bibfnamefont {M.}~\bibnamefont {Lein}},\ }\bibfield  {title} {\bibinfo {title} {High-order harmonic generation in the presence of a resonance},\ }\href {https://doi.org/10.1103/PhysRevA.84.013430} {\bibfield  {journal} {\bibinfo  {journal} {Physical Review A}\ }\textbf {\bibinfo {volume} {84}},\ \bibinfo {pages} {013430} (\bibinfo {year} {2011})},\ \bibinfo {note} {publisher: American Physical Society}\BibitemShut {NoStop}%
\bibitem [{\citenamefont {Ganeev}\ \emph {et~al.}(2012)\citenamefont {Ganeev}, \citenamefont {Witting}, \citenamefont {Hutchison}, \citenamefont {Frank}, \citenamefont {Tudorovskaya}, \citenamefont {Lein}, \citenamefont {Okell}, \citenamefont {Zaïr}, \citenamefont {Marangos},\ and\ \citenamefont {Tisch}}]{ganeev_isolated_2012}%
  \BibitemOpen
  \bibfield  {author} {\bibinfo {author} {\bibfnamefont {R.~A.}\ \bibnamefont {Ganeev}}, \bibinfo {author} {\bibfnamefont {T.}~\bibnamefont {Witting}}, \bibinfo {author} {\bibfnamefont {C.}~\bibnamefont {Hutchison}}, \bibinfo {author} {\bibfnamefont {F.}~\bibnamefont {Frank}}, \bibinfo {author} {\bibfnamefont {M.}~\bibnamefont {Tudorovskaya}}, \bibinfo {author} {\bibfnamefont {M.}~\bibnamefont {Lein}}, \bibinfo {author} {\bibfnamefont {W.~A.}\ \bibnamefont {Okell}}, \bibinfo {author} {\bibfnamefont {A.}~\bibnamefont {Zaïr}}, \bibinfo {author} {\bibfnamefont {J.~P.}\ \bibnamefont {Marangos}},\ and\ \bibinfo {author} {\bibfnamefont {J.~W.~G.}\ \bibnamefont {Tisch}},\ }\bibfield  {title} {\bibinfo {title} {Isolated sub-fs {XUV} pulse generation in {Mn} plasma ablation},\ }\href {https://doi.org/10.1364/OE.20.025239} {\bibfield  {journal} {\bibinfo  {journal} {Optics Express}\ }\textbf {\bibinfo {volume} {20}},\ \bibinfo {pages} {25239} (\bibinfo {year} {2012})}\BibitemShut {NoStop}%
\bibitem [{\citenamefont {Romanov}\ \emph {et~al.}(2024)\citenamefont {Romanov}, \citenamefont {Strelkov},\ and\ \citenamefont {Silaev}}]{romanov_simulation_2024}%
  \BibitemOpen
  \bibfield  {author} {\bibinfo {author} {\bibfnamefont {A.~A.}\ \bibnamefont {Romanov}}, \bibinfo {author} {\bibfnamefont {V.~V.}\ \bibnamefont {Strelkov}},\ and\ \bibinfo {author} {\bibfnamefont {A.~A.}\ \bibnamefont {Silaev}},\ }\bibfield  {title} {\bibinfo {title} {Simulation of resonance-enhanced high-order harmonic generation and autoionization decay in {Ga} + based on time-dependent density-functional theory},\ }\href {https://doi.org/10.1103/PhysRevA.110.063109} {\bibfield  {journal} {\bibinfo  {journal} {Physical Review A}\ }\textbf {\bibinfo {volume} {110}},\ \bibinfo {pages} {063109} (\bibinfo {year} {2024})}\BibitemShut {NoStop}%
\bibitem [{\citenamefont {Romanov}\ \emph {et~al.}(2021)\citenamefont {Romanov}, \citenamefont {Silaev}, \citenamefont {Sarantseva}, \citenamefont {Frolov},\ and\ \citenamefont {Vvedenskii}}]{romanov_study_2021}%
  \BibitemOpen
  \bibfield  {author} {\bibinfo {author} {\bibfnamefont {A.~A.}\ \bibnamefont {Romanov}}, \bibinfo {author} {\bibfnamefont {A.~A.}\ \bibnamefont {Silaev}}, \bibinfo {author} {\bibfnamefont {T.~S.}\ \bibnamefont {Sarantseva}}, \bibinfo {author} {\bibfnamefont {M.~V.}\ \bibnamefont {Frolov}},\ and\ \bibinfo {author} {\bibfnamefont {N.~V.}\ \bibnamefont {Vvedenskii}},\ }\bibfield  {title} {\bibinfo {title} {Study of high-order harmonic generation in xenon based on time-dependent density-functional theory},\ }\href {https://doi.org/10.1088/1367-2630/abe8a9} {\bibfield  {journal} {\bibinfo  {journal} {New Journal of Physics}\ }\textbf {\bibinfo {volume} {23}},\ \bibinfo {pages} {043014} (\bibinfo {year} {2021})},\ \bibinfo {note} {publisher: IOP Publishing}\BibitemShut {NoStop}%
\bibitem [{\citenamefont {Wahyutama}\ \emph {et~al.}(2019)\citenamefont {Wahyutama}, \citenamefont {Sato},\ and\ \citenamefont {Ishikawa}}]{wahyutama_time-dependent_2019}%
  \BibitemOpen
  \bibfield  {author} {\bibinfo {author} {\bibfnamefont {I.~S.}\ \bibnamefont {Wahyutama}}, \bibinfo {author} {\bibfnamefont {T.}~\bibnamefont {Sato}},\ and\ \bibinfo {author} {\bibfnamefont {K.~L.}\ \bibnamefont {Ishikawa}},\ }\bibfield  {title} {\bibinfo {title} {Time-dependent multiconfiguration self-consistent-field study on resonantly enhanced high-order harmonic generation from transition-metal elements},\ }\href {https://doi.org/10.1103/PhysRevA.99.063420} {\bibfield  {journal} {\bibinfo  {journal} {Physical Review A}\ }\textbf {\bibinfo {volume} {99}},\ \bibinfo {pages} {063420} (\bibinfo {year} {2019})}\BibitemShut {NoStop}%
\bibitem [{\citenamefont {Tikhomirov}\ \emph {et~al.}(2017)\citenamefont {Tikhomirov}, \citenamefont {Sato},\ and\ \citenamefont {Ishikawa}}]{tikhomirov_high-harmonic_2017}%
  \BibitemOpen
  \bibfield  {author} {\bibinfo {author} {\bibfnamefont {I.}~\bibnamefont {Tikhomirov}}, \bibinfo {author} {\bibfnamefont {T.}~\bibnamefont {Sato}},\ and\ \bibinfo {author} {\bibfnamefont {K.~L.}\ \bibnamefont {Ishikawa}},\ }\bibfield  {title} {\bibinfo {title} {High-{Harmonic} {Generation} {Enhanced} by {Dynamical} {Electron} {Correlation}},\ }\href {https://doi.org/10.1103/PhysRevLett.118.203202} {\bibfield  {journal} {\bibinfo  {journal} {Physical Review Letters}\ }\textbf {\bibinfo {volume} {118}},\ \bibinfo {pages} {203202} (\bibinfo {year} {2017})},\ \bibinfo {note} {publisher: American Physical Society}\BibitemShut {NoStop}%
\bibitem [{\citenamefont {Ganeev}\ \emph {et~al.}(2021)\citenamefont {Ganeev}, \citenamefont {Kuroda}, \citenamefont {Kuroda},\ and\ \citenamefont {Kuroda}}]{ganeev_resonance-affected_2021}%
  \BibitemOpen
  \bibfield  {author} {\bibinfo {author} {\bibfnamefont {R.~A.}\ \bibnamefont {Ganeev}}, \bibinfo {author} {\bibfnamefont {H.}~\bibnamefont {Kuroda}}, \bibinfo {author} {\bibfnamefont {H.}~\bibnamefont {Kuroda}},\ and\ \bibinfo {author} {\bibfnamefont {H.}~\bibnamefont {Kuroda}},\ }\bibfield  {title} {\bibinfo {title} {Resonance-affected high-order harmonic emission from atomic and molecular chromium laser-induced plasmas},\ }\href {https://doi.org/10.1364/OSAC.422269} {\bibfield  {journal} {\bibinfo  {journal} {OSA Continuum}\ }\textbf {\bibinfo {volume} {4}},\ \bibinfo {pages} {1545} (\bibinfo {year} {2021})},\ \bibinfo {note} {publisher: Optical Society of America}\BibitemShut {NoStop}%
\bibitem [{\citenamefont {Neufeld}\ \emph {et~al.}(2024{\natexlab{a}})\citenamefont {Neufeld}, \citenamefont {Tancogne-Dejean},\ and\ \citenamefont {Rubio}}]{neufeld_bench}%
  \BibitemOpen
  \bibfield  {author} {\bibinfo {author} {\bibfnamefont {O.}~\bibnamefont {Neufeld}}, \bibinfo {author} {\bibfnamefont {N.}~\bibnamefont {Tancogne-Dejean}},\ and\ \bibinfo {author} {\bibfnamefont {A.}~\bibnamefont {Rubio}},\ }\bibfield  {title} {\bibinfo {title} {Benchmarking functionals for strong-field light-matter interactions in adiabatic time-dependent density functional theory},\ }\href {https://doi.org/10.1021/acs.jpclett.4c01383} {\bibfield  {journal} {\bibinfo  {journal} {The Journal of Physical Chemistry Letters}\ }\textbf {\bibinfo {volume} {15}},\ \bibinfo {pages} {7254} (\bibinfo {year} {2024}{\natexlab{a}})},\ \bibinfo {note} {pMID: 38976844},\ \Eprint {https://arxiv.org/abs/https://doi.org/10.1021/acs.jpclett.4c01383} {https://doi.org/10.1021/acs.jpclett.4c01383} \BibitemShut {NoStop}%
\bibitem [{\citenamefont {Haessler}\ \emph {et~al.}(2013)\citenamefont {Haessler}, \citenamefont {Strelkov}, \citenamefont {Bom}, \citenamefont {Khokhlova}, \citenamefont {Gobert}, \citenamefont {Hergott}, \citenamefont {Lepetit}, \citenamefont {Perdrix}, \citenamefont {Ozaki},\ and\ \citenamefont {Sali{\`e}res}}]{haessler_phase_nodate}%
  \BibitemOpen
  \bibfield  {author} {\bibinfo {author} {\bibfnamefont {S.}~\bibnamefont {Haessler}}, \bibinfo {author} {\bibfnamefont {V.}~\bibnamefont {Strelkov}}, \bibinfo {author} {\bibfnamefont {L.~E.}\ \bibnamefont {Bom}}, \bibinfo {author} {\bibfnamefont {M.}~\bibnamefont {Khokhlova}}, \bibinfo {author} {\bibfnamefont {O.}~\bibnamefont {Gobert}}, \bibinfo {author} {\bibfnamefont {J.-F.}\ \bibnamefont {Hergott}}, \bibinfo {author} {\bibfnamefont {F.}~\bibnamefont {Lepetit}}, \bibinfo {author} {\bibfnamefont {M.}~\bibnamefont {Perdrix}}, \bibinfo {author} {\bibfnamefont {T.}~\bibnamefont {Ozaki}},\ and\ \bibinfo {author} {\bibfnamefont {P.}~\bibnamefont {Sali{\`e}res}},\ }\bibfield  {title} {\bibinfo {title} {Phase distortions of attosecond pulses produced by resonance-enhanced high harmonic generation},\ }\href@noop {} {\bibfield  {journal} {\bibinfo  {journal} {New Journal of Physics}\ }\textbf {\bibinfo {volume} {15}},\ \bibinfo {pages} {013051} (\bibinfo {year} {2013})}\BibitemShut {NoStop}%
\bibitem [{\citenamefont {Sansonetti}\ and\ \citenamefont {Nave}(2014)}]{sansonetti2014extended}%
  \BibitemOpen
  \bibfield  {author} {\bibinfo {author} {\bibfnamefont {C.~J.}\ \bibnamefont {Sansonetti}}\ and\ \bibinfo {author} {\bibfnamefont {G.}~\bibnamefont {Nave}},\ }\bibfield  {title} {\bibinfo {title} {Extended analysis of the spectrum of singly ionized chromium (cr ii)},\ }\href@noop {} {\bibfield  {journal} {\bibinfo  {journal} {The Astrophysical Journal Supplement Series}\ }\textbf {\bibinfo {volume} {213}},\ \bibinfo {pages} {28} (\bibinfo {year} {2014})}\BibitemShut {NoStop}%
\bibitem [{\citenamefont {Moiseyev}(1998)}]{moiseyev_quantum_1998}%
  \BibitemOpen
  \bibfield  {author} {\bibinfo {author} {\bibfnamefont {N.}~\bibnamefont {Moiseyev}},\ }\bibfield  {title} {\bibinfo {title} {Quantum theory of resonances: calculating energies, widths and cross-sections by complex scaling},\ }\href {https://doi.org/10.1016/S0370-1573(98)00002-7} {\bibfield  {journal} {\bibinfo  {journal} {Physics Reports}\ }\textbf {\bibinfo {volume} {302}},\ \bibinfo {pages} {212} (\bibinfo {year} {1998})}\BibitemShut {NoStop}%
\bibitem [{\citenamefont {Moiseyev}(2011)}]{Moiseyev_2011}%
  \BibitemOpen
  \bibfield  {author} {\bibinfo {author} {\bibfnamefont {N.}~\bibnamefont {Moiseyev}},\ }\href@noop {} {\emph {\bibinfo {title} {Non-Hermitian Quantum Mechanics}}}\ (\bibinfo  {publisher} {Cambridge University Press},\ \bibinfo {year} {2011})\BibitemShut {NoStop}%
\bibitem [{\citenamefont {Neufeld}\ and\ \citenamefont {Cohen}(2019)}]{neufeld_background-free_2019}%
  \BibitemOpen
  \bibfield  {author} {\bibinfo {author} {\bibfnamefont {O.}~\bibnamefont {Neufeld}}\ and\ \bibinfo {author} {\bibfnamefont {O.}~\bibnamefont {Cohen}},\ }\bibfield  {title} {\bibinfo {title} {Background-{Free} {Measurement} of {Ring} {Currents} by {Symmetry}-{Breaking} {High}-{Harmonic} {Spectroscopy}},\ }\href {https://doi.org/10.1103/PhysRevLett.123.103202} {\bibfield  {journal} {\bibinfo  {journal} {Physical Review Letters}\ }\textbf {\bibinfo {volume} {123}},\ \bibinfo {pages} {103202} (\bibinfo {year} {2019})},\ \bibinfo {note} {publisher: American Physical Society}\BibitemShut {NoStop}%
\bibitem [{\citenamefont {Johnson}(2021)}]{johnson2021notesperfectlymatchedlayers}%
  \BibitemOpen
  \bibfield  {author} {\bibinfo {author} {\bibfnamefont {S.~G.}\ \bibnamefont {Johnson}},\ }\href {https://arxiv.org/abs/2108.05348} {\bibinfo {title} {Notes on perfectly matched layers (pmls)}} (\bibinfo {year} {2021}),\ \Eprint {https://arxiv.org/abs/2108.05348} {arXiv:2108.05348 [cs.CE]} \BibitemShut {NoStop}%
\bibitem [{\citenamefont {Miller}(2016)}]{miller2016}%
  \BibitemOpen
  \bibfield  {author} {\bibinfo {author} {\bibfnamefont {M.~R.}\ \bibnamefont {Miller}},\ }\href@noop {} {} (\bibinfo {year} {2016}),\ \bibinfo {note} {\emph{Time Resolving Electron Dynamics in Atomic and Molecular Systems Using High-Harmonic Spectroscopy}, {Ph.D.} thesis, University of Colorado Boulder}\BibitemShut {NoStop}%
\bibitem [{\citenamefont {Marques}\ and\ \citenamefont {Gross}(2004)}]{marques_time-dependent_2004}%
  \BibitemOpen
  \bibfield  {author} {\bibinfo {author} {\bibfnamefont {M.~a.~L.}\ \bibnamefont {Marques}}\ and\ \bibinfo {author} {\bibfnamefont {E.~K.~U.}\ \bibnamefont {Gross}},\ }\bibfield  {title} {\bibinfo {title} {Time-dependent density functional theory},\ }\href {https://doi.org/10.1146/annurev.physchem.55.091602.094449} {\bibfield  {journal} {\bibinfo  {journal} {Annual Review of Physical Chemistry}\ }\textbf {\bibinfo {volume} {55}},\ \bibinfo {pages} {427} (\bibinfo {year} {2004})},\ \bibinfo {note} {publisher: Annual Reviews}\BibitemShut {NoStop}%
\bibitem [{\citenamefont {Tancogne-Dejean}\ \emph {et~al.}(2020)\citenamefont {Tancogne-Dejean}, \citenamefont {Oliveira}, \citenamefont {Andrade}, \citenamefont {Appel}, \citenamefont {Borca}, \citenamefont {Le~Breton}, \citenamefont {Buchholz}, \citenamefont {Castro}, \citenamefont {Corni}, \citenamefont {Correa}, \citenamefont {De~Giovannini}, \citenamefont {Delgado}, \citenamefont {Eich}, \citenamefont {Flick}, \citenamefont {Gil}, \citenamefont {Gomez}, \citenamefont {Helbig}, \citenamefont {Hübener}, \citenamefont {Jestädt}, \citenamefont {Jornet-Somoza}, \citenamefont {Larsen}, \citenamefont {Lebedeva}, \citenamefont {Lüders}, \citenamefont {Marques}, \citenamefont {Ohlmann}, \citenamefont {Pipolo}, \citenamefont {Rampp}, \citenamefont {Rozzi}, \citenamefont {Strubbe}, \citenamefont {Sato}, \citenamefont {Schäfer}, \citenamefont {Theophilou}, \citenamefont {Welden},\ and\ \citenamefont {Rubio}}]{tancogne-dejean_octopus_2020}%
  \BibitemOpen
  \bibfield  {author} {\bibinfo {author} {\bibfnamefont {N.}~\bibnamefont {Tancogne-Dejean}}, \bibinfo {author} {\bibfnamefont {M.~J.~T.}\ \bibnamefont {Oliveira}}, \bibinfo {author} {\bibfnamefont {X.}~\bibnamefont {Andrade}}, \bibinfo {author} {\bibfnamefont {H.}~\bibnamefont {Appel}}, \bibinfo {author} {\bibfnamefont {C.~H.}\ \bibnamefont {Borca}}, \bibinfo {author} {\bibfnamefont {G.}~\bibnamefont {Le~Breton}}, \bibinfo {author} {\bibfnamefont {F.}~\bibnamefont {Buchholz}}, \bibinfo {author} {\bibfnamefont {A.}~\bibnamefont {Castro}}, \bibinfo {author} {\bibfnamefont {S.}~\bibnamefont {Corni}}, \bibinfo {author} {\bibfnamefont {A.~A.}\ \bibnamefont {Correa}}, \bibinfo {author} {\bibfnamefont {U.}~\bibnamefont {De~Giovannini}}, \bibinfo {author} {\bibfnamefont {A.}~\bibnamefont {Delgado}}, \bibinfo {author} {\bibfnamefont {F.~G.}\ \bibnamefont {Eich}}, \bibinfo {author} {\bibfnamefont {J.}~\bibnamefont {Flick}}, \bibinfo {author} {\bibfnamefont {G.}~\bibnamefont {Gil}}, \bibinfo {author} {\bibfnamefont
  {A.}~\bibnamefont {Gomez}}, \bibinfo {author} {\bibfnamefont {N.}~\bibnamefont {Helbig}}, \bibinfo {author} {\bibfnamefont {H.}~\bibnamefont {Hübener}}, \bibinfo {author} {\bibfnamefont {R.}~\bibnamefont {Jestädt}}, \bibinfo {author} {\bibfnamefont {J.}~\bibnamefont {Jornet-Somoza}}, \bibinfo {author} {\bibfnamefont {A.~H.}\ \bibnamefont {Larsen}}, \bibinfo {author} {\bibfnamefont {I.~V.}\ \bibnamefont {Lebedeva}}, \bibinfo {author} {\bibfnamefont {M.}~\bibnamefont {Lüders}}, \bibinfo {author} {\bibfnamefont {M.~A.~L.}\ \bibnamefont {Marques}}, \bibinfo {author} {\bibfnamefont {S.~T.}\ \bibnamefont {Ohlmann}}, \bibinfo {author} {\bibfnamefont {S.}~\bibnamefont {Pipolo}}, \bibinfo {author} {\bibfnamefont {M.}~\bibnamefont {Rampp}}, \bibinfo {author} {\bibfnamefont {C.~A.}\ \bibnamefont {Rozzi}}, \bibinfo {author} {\bibfnamefont {D.~A.}\ \bibnamefont {Strubbe}}, \bibinfo {author} {\bibfnamefont {S.~A.}\ \bibnamefont {Sato}}, \bibinfo {author} {\bibfnamefont {C.}~\bibnamefont {Schäfer}}, \bibinfo {author}
  {\bibfnamefont {I.}~\bibnamefont {Theophilou}}, \bibinfo {author} {\bibfnamefont {A.}~\bibnamefont {Welden}},\ and\ \bibinfo {author} {\bibfnamefont {A.}~\bibnamefont {Rubio}},\ }\bibfield  {title} {\bibinfo {title} {Octopus, a computational framework for exploring light-driven phenomena and quantum dynamics in extended and finite systems},\ }\href {https://doi.org/10.1063/1.5142502} {\bibfield  {journal} {\bibinfo  {journal} {The Journal of Chemical Physics}\ }\textbf {\bibinfo {volume} {152}},\ \bibinfo {pages} {124119} (\bibinfo {year} {2020})}\BibitemShut {NoStop}%
\bibitem [{\citenamefont {Legrand}\ \emph {et~al.}(2002)\citenamefont {Legrand}, \citenamefont {Suraud},\ and\ \citenamefont {Reinhard}}]{CLegrand_2002}%
  \BibitemOpen
  \bibfield  {author} {\bibinfo {author} {\bibfnamefont {C.}~\bibnamefont {Legrand}}, \bibinfo {author} {\bibfnamefont {E.}~\bibnamefont {Suraud}},\ and\ \bibinfo {author} {\bibfnamefont {P.-G.}\ \bibnamefont {Reinhard}},\ }\bibfield  {title} {\bibinfo {title} {Comparison of self-interaction-corrections for metal clusters},\ }\href {https://doi.org/10.1088/0953-4075/35/4/333} {\bibfield  {journal} {\bibinfo  {journal} {Journal of Physics B: Atomic, Molecular and Optical Physics}\ }\textbf {\bibinfo {volume} {35}},\ \bibinfo {pages} {1115} (\bibinfo {year} {2002})}\BibitemShut {NoStop}%
\bibitem [{\citenamefont {Neufeld}\ \emph {et~al.}(2024{\natexlab{b}})\citenamefont {Neufeld}, \citenamefont {Tancogne-Dejean},\ and\ \citenamefont {Rubio}}]{neufeld_benchmarking_2024}%
  \BibitemOpen
  \bibfield  {author} {\bibinfo {author} {\bibfnamefont {O.}~\bibnamefont {Neufeld}}, \bibinfo {author} {\bibfnamefont {N.}~\bibnamefont {Tancogne-Dejean}},\ and\ \bibinfo {author} {\bibfnamefont {A.}~\bibnamefont {Rubio}},\ }\bibfield  {title} {\bibinfo {title} {Benchmarking {Functionals} for {Strong}-{Field} {Light}-{Matter} {Interactions} in {Adiabatic} {Time}-{Dependent} {Density} {Functional} {Theory}},\ }\href {https://doi.org/10.1021/acs.jpclett.4c01383} {\bibfield  {journal} {\bibinfo  {journal} {The Journal of Physical Chemistry Letters}\ }\textbf {\bibinfo {volume} {15}},\ \bibinfo {pages} {7254} (\bibinfo {year} {2024}{\natexlab{b}})},\ \bibinfo {note} {publisher: American Chemical Society}\BibitemShut {NoStop}%
\bibitem [{\citenamefont {Hartwigsen}\ \emph {et~al.}(1998)\citenamefont {Hartwigsen}, \citenamefont {Goedecker},\ and\ \citenamefont {Hutter}}]{hartwigsen_relativistic_1998}%
  \BibitemOpen
  \bibfield  {author} {\bibinfo {author} {\bibfnamefont {C.}~\bibnamefont {Hartwigsen}}, \bibinfo {author} {\bibfnamefont {S.}~\bibnamefont {Goedecker}},\ and\ \bibinfo {author} {\bibfnamefont {J.}~\bibnamefont {Hutter}},\ }\bibfield  {title} {\bibinfo {title} {Relativistic separable dual-space {Gaussian} pseudopotentials from {H} to {Rn}},\ }\href {https://doi.org/10.1103/PhysRevB.58.3641} {\bibfield  {journal} {\bibinfo  {journal} {Physical Review B}\ }\textbf {\bibinfo {volume} {58}},\ \bibinfo {pages} {3641} (\bibinfo {year} {1998})},\ \bibinfo {note} {publisher: American Physical Society}\BibitemShut {NoStop}%
\bibitem [{\citenamefont {Bray}\ \emph {et~al.}(2020)\citenamefont {Bray}, \citenamefont {Freeman}, \citenamefont {Naseem}, \citenamefont {Dolmatov},\ and\ \citenamefont {Kheifets}}]{bray_correlation-enhanced_2020}%
  \BibitemOpen
  \bibfield  {author} {\bibinfo {author} {\bibfnamefont {A.~W.}\ \bibnamefont {Bray}}, \bibinfo {author} {\bibfnamefont {D.}~\bibnamefont {Freeman}}, \bibinfo {author} {\bibfnamefont {F.}~\bibnamefont {Naseem}}, \bibinfo {author} {\bibfnamefont {V.~K.}\ \bibnamefont {Dolmatov}},\ and\ \bibinfo {author} {\bibfnamefont {A.~S.}\ \bibnamefont {Kheifets}},\ }\bibfield  {title} {\bibinfo {title} {Correlation-enhanced high-order-harmonic-generation spectra of {Mn} and {Mn} +},\ }\href {https://doi.org/10.1103/PhysRevA.101.053415} {\bibfield  {journal} {\bibinfo  {journal} {Physical Review A}\ }\textbf {\bibinfo {volume} {101}},\ \bibinfo {pages} {053415} (\bibinfo {year} {2020})}\BibitemShut {NoStop}%
\end{thebibliography}%
\newpage

\section*{Computational Methods}  \label{sec:Methodology}
We employed three theoretical descriptions of rHHG from Cr$^{+}$. First, we implemented a 1D scheme with a potential supporting a shape-resonance (metastable state, MS). We used a square barrier-well system, smoothing the domain edges using a hyperbolic function $\tanh(\alpha x)$, with $\alpha =\frac{1}{8 \, \text{a.u.}}$ (see  SI):
\( V(x) = V_w \,\, \text{if } |x| \le l_w;\quad V_b \,\, \text{if } l_w < |x| \le l_w + l_b;\quad 0 \,\, \text{otherwise} \), 
where $l_w$ and $l_b$ are the widths of the potential well and barrier respectively, $V_w<0$ is depth of the potential well and $V_b>0$ is the barrier height. 
The values of the ionization potential $I_p$ and the MS energy $E_{MS}$ approximately match that of Cr$^{+}$ (model $I_p\sim$15.8 eV, experimental $I_p\sim$16.4 eV\cite{sansonetti2014extended}). The energy levels, including $I_p$ and $E_{MS}$, are computed using complex-scaling \cite{moiseyev_quantum_1998}. Tuning the model parameters allows to systematically control the lifetime of the MS state, e.g. by increasing the barrier width (reducing the tunneling rate), thereby increasing the lifetime, while altering the ground-state energy and wavefunction only negligibly. 
Following ref. \cite{Moiseyev_2011}, the lifetime is calculated as \begin{math}
    \tau=-\frac{\hbar}{2\text{Im}\left[E_{res}\right]}
\end{math}
The time-dependent Schr{\"o}dinger equation (TDSE) is represented on a real-space grid where the initial state is an $s$-like (symmetric) ground state and the shape resonance residing in a $p$-like level.
The driving field is of the form $E(t)=f(t) \cos(\omega t)$, where $f(t)$ is an envelope function given by a super-sine function \cite{neufeld_background-free_2019}, remains constant, and then decays. The driving frequency $\omega$ is set such that H29 is in resonance with the ground-MS transition $29\omega= I_p + E_{MS}$. For the main simulation, corresponding to a MS lifetime of 0.17T, this results in $\omega=0.0621$ [a.u.], or $734$ nm. We solved the TDSE in the dipole approximation and length gauge, given in atomic units by: 
\begin{equation}
\left[i\frac{\partial}{\partial t}+\frac{1}{2}\frac{\partial^{2}}{\partial x^{2}}-V\left(x\right)-xE\left(t\right)\right]\psi\left(x\right)=0
\end{equation}
and implement CAP in the form of perfectly-matched layers (PML) \cite{johnson2021notesperfectlymatchedlayers}. 

From the propagated state, $\psi(x,t)$, we obtain the dipole acceleration through Ehrenfest's theorem \cite{miller2016} as $a(t)=\frac{1}{i}\left\langle \left[\hat{P},\hat{H}\right]\right\rangle $,and calculate the HHG spectrum by taking the Fourier transform: $I(\Omega)=\left| \frac{1}{\sqrt{2 \pi}}\intop\limits_{-\infty}^{\infty}a(t)e^{-i\Omega t}dt \right |^2$. Time domain information is obtained with a Gabor analysis (see SI). 

In the second approach, we employed state-of-the-art time-dependent spin density functional theory (TDDFT) \cite {marques_time-dependent_2004} using Octopus code \cite{tancogne-dejean_octopus_2020}. This approach is fully \textit{ab-initio} (in 3D). The ground state of Cr$^{+}$ is calculated first with DFT (utilizing the local spin density approximation with an added self-interaction correction,\cite{CLegrand_2002} which performs well for HHG\cite{neufeld_benchmarking_2024}). In the ground state, the system is magnetic with half-filled $3d$ orbitals, while all deeper levels are fully occupied (see Fig. \ref{fig:TDFT_gabor}(a)). Semi-core $3p$ levels are included in the calculation, while deeper states are treated by norm-conserving pseudopotentials \cite{hartwigsen_relativistic_1998}. After obtaining the ground state, the system's response to an 800 nm laser pulse with similar shape and intensity as in the 1D case is calculated by solving the time-dependent Kohn-Sham (KS) equations

\begin{equation} \label{eq:KS_Schrodinger}
i\partial_t\ \psi_{n}^{KS}(\textbf{r},t) = \left[-\frac{1}{2}\nabla^2 + v_{KS}(\textbf{r}) - \textbf{E}(t)\cdot \textbf{r} \right] \psi_{n}^{KS}(\textbf{r},t) ,
\end{equation}

where $n$ is the orbital index, $\textbf{r}$ is the electronic coordinate, $\psi_{n}^{KS}(\textbf{r},t)$ are the time-dependent KS orbitals, and $v_{KS}$ is the KS potential:

\begin{equation}\label{eq:KS_potential}
v_{KS}(\textbf{r},t) = v_{e-ion} + \int d^3r'\frac{n(\mathbf{r'},t)}{|\mathbf{r}-\mathbf{r'}|}+v_{XC}[\rho(\mathbf{r},t)] ,
\end{equation}
where the first term in eq. \ref{eq:KS_potential} describes electron-ion interactions (including also interactions with the frozen core states through the pseudopotentials, as well as a self-interaction correction), the second term is the electrostatic Hartree repulsive interaction between electrons, and the last term is the exchange-correlation potential taken in the adiabatic and local spin density approximation with $\rho(\textbf{r},t)$ the spin density matrix (from which the electron density $n(\textbf{r},t)$ can be obtained). In this notation the orbital index $n$ also refers to spin sub-levels, since up/down spin channels are no longer energy-degenerate (e.g. as a result of spin exchange interactions), though the actual deviations are small. 

The TDDFT equations for all initially occupied levels are propagated in real-time (see SI for further details), from which we obtain the HHG spectra and time-frequency analysis in a similar manner to the 1D model, but with separation into spin channels (by separating up/down channel contribution to $n(\textbf{r},t)$). Time-dependent simulations include a CAP.

We employ a third technique where the KS equations are solved by applying the independent particle approximation (IPA), i.e, fixing the KS potential to its form at $t=0$ with $v_{ks}[\rho(\textbf{r},t)]=v_{ks}[\rho(\textbf{r},t=0)]$. The different orbitals decouple and no longer interact through mean-field terms, only through the effective potential of the ground-state system, removing dynamic \textit{e-e} interactions. The IPA is usually a very good approximation in strong-field physics; however, it has already been shown to be inadequate in resonance HHG from Ga ions \cite{romanov_simulation_2024}, and is not expected to hold also in Cr$^{+}$ \cite{wahyutama_time-dependent_2019, bray_correlation-enhanced_2020}.

\end{document}


\begin{center}
\textbf{\large Supplementary Information for \\[6pt]
“Correlation-induced phase shifts and time delays in resonance enhanced high harmonic generation from Cr$^{+}$”}
\end{center}

\title{Correlation-induced phase shifts and time delays in resonance enhanced high harmonic generation from Cr$^{+}$}

\author{Yoad Aharon}
\author{Adi Pick}
\author{Amir Hen}
\author{Gilad Marcus}
\email{gilad.marcus@mail.huji.ac.il}
\affiliation{Institute of Applied Physics, Hebrew University of Jerusalem, Jerusalem 9190401, Israel}

\author{Ofer Neufeld}
\email{ofern@technion.ac.il}
\affiliation{Technion Israel Institute of Technology, Faculty of Chemistry, Haifa 3200003, Israel}

\date{\today}

\section{Additional technical details}
\subsection{1D model}
\subsubsection{The 1D potential}
To construct a rounded potential, we first defined the function:
\begin{equation}
u\left(x;l\right)=\frac{\tanh\left(\alpha\left(x-\frac{l}{2}\right)\right)-\tanh\left(\alpha\left(x+\frac{l}{2}\right)\right)}{2}
\end{equation}
which constitutes a rounded analogue of the function $-\text{rect}\left( \frac{x}{l} \right)$. Here,  the parameter $\alpha$ controls the degree of smoothing:  
\begin{equation}
    \lim_{\alpha\rightarrow \infty} u\left(x;l\right)=-\text{rect}\left( \frac{x}{l} \right)
\end{equation}
This function served as a building block for the barriers and the well:
\begin{equation}
V(x)=\underset{\text{well}}{\underbrace{V_{w}\,u\left(x;l_{w}\right)}}+
\underset{\text{barriers}}{\underbrace{V_{b}\left[u\left(x+\frac{l_{w}}{2};l_{b}\right)+u\left(x-\frac{l_{w}}{2};l_{b}\right)\right]}}
\end{equation}
\subsubsection{Hamiltonian Construction}
The one dimensional (1D) model Hamiltonian was constructed on a real-space grid with the kinetic energy term represented using a second-order central-difference approximation for the second derivative. The box size was chosen as $L = 100$ (a.u.), with step size $\Delta x = 0.28$ (a.u). The main potential parameters discussed in the main text define the well-barrier system and resulting resonance state: well-depth $V_w$ , well width $l_w$, barrier height $V_b$, and barrier width $l_b$. 
The resonance energy was tuned by varying each of the four parameters of the potential (calculations discussed below). 
For all the 1D simulations we chose the following values (in [a.u.]: $l_w=1.7$; $V_w=-1.4$; $V_b=1.5$; $\alpha=8$. For our main simulation, with an MS lifetime of $0.17T$, we chose $l_b=2$.

For time-dependent simulations the TDSE was solved with a variable time step using MATLAB's \cite{MATLAB2023} native ode45 solver, and with a laser wavelength of $\lambda = 800 \, \text{nm}$ and a typical laser power of  $3.51\times10^{14} \, W/cm^2$, and pulse duration of $21.23\,fs$ full width half max (FWHM). Under these parameters the resonant enhancement arises in the 29th harmonic. To prevent reflections from the boundaries of the box, we employed the method of perfectly matched layers (PML) \cite{johnson2021notesperfectlymatchedlayers}. The width of the PML was set to one-fifth of the box size on each side.

\subsubsection{Complex-Scaling}
To study the resonance enhancement of HHG, knowledge and control of the energy and life-time of the meta-stable state are crucial. For a given potential exhibiting a shape-resonance, this information can be numerically calculated by means of the complex-scaling method \cite{moiseyev_quantum_1998}. 

In complex scaling, the position operator undergoes an analytical continuation defined as $\hat{X}(\theta) = e^{i\theta}\hat{X}$. Consequently, the momentum operator scales as $\hat{P}(\theta) = e^{-i\theta}\hat{P}$, leading to the following scaling of the Hamiltonian:
$
\hat{H}(\theta) = \frac{e^{-2i\theta}\hat{P}^2}{2} + V\left(e^{i\theta}\hat{X}\right)$.  It has been demonstrated \cite{aguilar_class_1971} that the negative eigenvalues of the Hamiltonian, corresponding to bound-state energies, remain unchanged under this transformation. In contrast, the continuum-state energies scale as $E(\theta) = |E|e^{-2i\theta}$.  For meta-stable states, commonly referred to as resonances, a critical angle $\theta_c$ exists. Beyond this angle the resonance energies cease to rotate, and assume to complex value $E(\theta_c) = \epsilon - i\Gamma$, where $\epsilon$ represents the resonance peak center, and $\Gamma$ denotes its width. 

\subsubsection{HHG Calculation}
The HHG spectrum was calculated by taking the Fourier transform of the dipole acceleration, which was derived using Ehrenfest theorem \cite{miller2016}
\begin{equation}
      \begin{split}
          \ddot{d}(t) &=  \frac{d}{dt}\langle\hat{P}\rangle=-\langle\frac{\partial}{\partial x}V\left(x,t\right)\rangle
          \\
        &=-\intop_{-L/2}^{L/2} \psi^\ast\left(x,t\right) \frac{\partial}{\partial x}V \left( x,t \right) \psi \left(x,t\right)dx
      \end{split}
\end{equation}
where $V\left(x,t\right)$ represents the sum of the atomic potential and the laser potential. The dipole acceleration was multiplied by a Blackman filter, and then zero padded on either side with 4 times the signal length, prior to taking the Fourier transform. To evaluate the Fourier transform we integrated over the entire zero-padded signal, using MATLAB's \cite{MATLAB2023} fft.
The Gabor transform was calculated by taking the Fourier transform of the dipole acceleration multiplied by a moving Gaussian window. The width of the window, defined by 1 standard deviation, was set to $\frac{1}{35}$ of the laser period, with an overlap of 95\% between each iteration. 
\begin{equation}
 GT\left(\tau\right)=\intop_{0}^{t_{fin}}\ddot{d} \, e^{-\frac{1}{2}\left(\frac{t-\tau}{\Delta t}\right)^{2}}e^{-i\omega t}dt
\end{equation}
This provides a way to analyze the spectral composition within a specific temporal window, independently of the remainder of the signal. By examining the dipole oscillations in this manner, we obtain the spectrum as a function of time. In the context of HHG, this reflects the recombination times of the different trajectories, each emitting radiation according to its kinetic energy.

\subsubsection{Group Delay Correction}
The group delay (GD) of a pulse is defined as \cite{weiner2011ultrafast}:
\begin{equation}
\tau_g \left( \omega \right)=-\frac{\partial \phi\left(\omega\right)}{\partial \omega}
\end{equation}
Note that here $\omega$ is treated as an independent variable, in contrast to eq. \ref{eqn:RABBIT_expression}, where $\omega$ denotes the driving frequency.
If a constant group delay $\tau_0$ is introduced, the phase is modified as:
\begin{equation}
    \tilde{\phi}=\phi-\omega\tau_0 
\end{equation}
such that
\begin{equation}
\begin{split}
        \tilde{\tau}&=-\frac{\partial \tilde{\phi}\left(\omega\right)}{\partial \omega}
        \\
        &=-\frac{\partial \phi\left(\omega\right)}{\partial \omega}+\tau_0
        \\
        &=\tau_g+\tau_0
\end{split}
\end{equation}
This is tantamount to a temporal shift, as would occur when traversing a material with a linear dispersion. This is evident from the shift property of the Fourier transform:
\begin{equation}
    \mathcal{F}\left[ E\left( t-\tau_0 \right)  \right] = \left|\tilde{E}\left(\omega\right)\right|e^{i \left( \phi(\omega) -\omega\tau_0 \right)}
\end{equation}
We may consider the group delay of an attosecond pulse, where the phases are extracted via a RABBIT measurement, as given by eq. \ref{eqn:SB_phi}. A temporal shift $\tau_0$ between the laser and the signal results in a corresponding shift in the argument of the SB cosine function from eq. \ref{eqn:RABBIT_SB}:
\begin{equation}\label{eqn:SB_GD}
    2\omega(\tau-\tau_0)+\Phi
\end{equation}
Note that in eq. \ref{eqn:SB_phi} $\omega$ refers to the driving frequency, whereas in the above discussion $\omega$ denotes the frequency as an independent variable in the spectral domain. 
The above may be written as:
\begin{equation}
    2\omega \tau +\tilde{\Phi}
\end{equation}
with $\tilde{\Phi}$  equal to:
\begin{equation}
    \tilde{\Phi}=\Phi -2\omega\tau_0
\end{equation}
In Fig. 4 (a) the harmonic phases due to the different models are compared. The 1D results (blue curve) shows the 1D phases with an addition of a constant group delay. We found a group delay of $-\frac{T}{8}$ aligns the 1D curve with that of the IPA model (orange curve). This is equivalent to an increase of $\frac{\pi}{4}$
between two adjacent (odd) harmonics.
 
\subsection{TDDFT simulations}
This section describes additional technical details about the DFT and TDDFT simulations employed throughout the main text. All DFT and TDDFT simulations were performed within Octopus code on a real-space grid, where the grid spacing was taken as 0.35 Bohr, and the grid was cartesian with spherical boundaries with a radius of 45 Bohr. Ground state simulations within spin DFT were first performed to obtain the initial KS orbitals (at $t=0$), where we employed the local density approximation with an added self-interaction correction as discussed in the main text. These states were then propagated within TDDFT using a time step of 2.9 attoseconds. At each time step we computed the induced dipole moment in the system (separated to spin-up and down channels), from which we numerically obtained the dipole acceleration and corresponding HHG spectra as described in the main text, but directly from the calculated dipoles (instead of employing Ehrenfest theorem), and where HHG spectra were also filtered with a temporal gaussian window. During propagation, we added a complex absorbing potential along the boundaries of width 15 Bohr to avoid reflections. 

IPA simulations were performed within the same methodology, except that the KS potential was frozen to its form at $t=0$, as discussed in the main text. 

Absorption and photoionization cross-section simulations presented in this SI were performed in a similar strategy, but by varying the driving laser wavelength and calculating: (i) The induced response at the driving wavelength (absorption), (ii) the total photo-ionization yield obtained via the portion of the electron density that was absorbed at the boundaries at the end of the simulation. 

Gabor transforms were calculated similarly to the 1D model with a gaussian window function with a width of a third of an optical period. 

\section{RABBIT analysis}
We performed a RABBIT-like calculation where the signal is expressed as:
\begin{equation}\label{eqn:RABBIT_expression}
    S= \left\{ \mathcal{F}\left[\left|\ddot{d}(t)+\cos\left(\omega\left(t-\tau\right)\right)\right|^{2}\right] \right\}^2
\end{equation}
where $\ddot{d}$ represents the second time derivative of the dipole moment's expectation value, and $\tau$ denotes the time delay between the local oscillator $\cos \left( \omega t \right)$ and $\ddot{d}(t)$. 
To demonstrate that this indeed retrieves the RABBIT spectrum, we first evaluate the argument of the Fourier transform:
\begin{equation}\label{eq:RABBIT_sig}
    I\left(t;\tau\right)=\left|\ddot{d}(t)+\cos\left(\omega\left(t-\tau\right)\right)\right|^{2}
\end{equation}
We may approximate the harmonic emission by the sum:
\begin{equation}\label{eqn:dipole_acceleration}
    \ddot{d}(t) =\sum_{q\in \text{odd}} c_{q}\cos\left(q \omega  t+\phi_{q}\right)
\end{equation}
where $\phi_q$ is the harmonic phase.
Substituting this sum into eq. \ref{eq:RABBIT_sig} we get:
\begin{equation}
    I\left(t;\tau\right)=\left|\sum_{q\in \text{odd}} c_{q} \cos\left(q \omega  t+\phi_{q}\right)+\cos\left(\omega\left(t-\tau\right)\right)\right|^{2}
\end{equation}
The expansion of the square in the above expression includes the following terms:
\begin{equation}
    I\propto2\left(c_{q}\cos\left(q\omega t+\phi_{q}\right)\cos\left(\omega\left(t-\tau\right)\right)+c_{q+2}\cos\left(\left(q+2\right)\omega t+\phi_{q+2}\right)\cos\left(\omega\left(t-\tau\right)\right)\right)
\end{equation}
Using the cosine sum-angle formula the above products simplify to the following sums:
\begin{equation}
    \begin{split}
  I\propto  2c_{q}\left[\cos\left(\left(q+1\right)\omega t-\omega\tau+\phi_{q}\right)+\cos\left(\left(q-1\right)\omega t+\omega\tau+\phi_{q}\right)\right] 
    \\
    + 2c_{q+2}\left[\cos\left(\left(q+3\right)\omega t-\omega\tau+\phi_{q+2}\right)+\cos\left(\left(q+1\right)\omega t+\omega\tau+\phi_{q+2}\right)\right]
\end{split}
\end{equation}
Since we are interested in the SB, we further investigate the terms with frequency $\left(q+1\right)\omega$:
\begin{equation}\label {eq:q+1 terms}
\begin{split}
        I_{q+1} &=2\left[c_{q}\cos\left(\left(q+1\right)\omega t-\omega\tau+\phi_{q}\right)+c_{q+2}\cos\left(\left(q+1\right)\omega t+\omega\tau+\phi_{q+2}\right)\right]
    \\
   & =e^{i\left(q+1\right)\omega t}\left(c_{q}e^{i\left(-\omega\tau+\phi_{q}\right)}+c_{q+2}e^{i\left(\omega\tau+\phi_{q+2}\right)}\right)+c.c.
\end{split}
    \end{equation}
Lastly, we evaluate the Fourier transform of eq.\ref{eq:q+1 terms} to obtain the $\left( q+1 \right) ^{th}$ SB:
\begin{equation}\label{eqn:RABBIT_SB}
\begin{split}
    \left|\tilde{F}\left(\left(q+1\right)\omega\right)\right|^{2} & \propto \left|c_{q}e^{i\left(-\omega\tau+\phi_{q}\right)}+c_{q+2}e^{i\left(\omega\tau+\phi_{q+2}\right)}\right|^{2}
    \\
    & =c_{q}^{2}+c_{q+2}^{2}+2c_{q}c_{q+2}\cos\left(2\omega\tau+\Phi_{q+1}\right)
    \end{split}
\end{equation}
where the SB phase $\Phi$  is equal to:
\begin{equation}
\begin{split}
   \Phi_{2}&=\phi_{3}-\phi_{1}
\\
\Phi_{4}&=\phi_{5}-\phi_{3}
\\
\Phi_{6}&=\phi_{7}-\phi_{5}
\\
&\qquad\vdots
\\
\Phi_{q+1}&=\phi_{q+2}-\phi_q
\end{split}
\end{equation}
The sideband signal (eq. \ref{eqn:RABBIT_SB}) oscillates as a function of delay $\tau$ with frequency $2\omega $, and has a phase equal to the difference between the phases of two adjacent harmonics. 
By summing over the SB phases we obtain the harmonic phase, up to a global phase:
\begin{equation}
\begin{split}
   \Phi_{2}+\Phi_{4}&=\cancel{\phi_3} - \phi_1 + \phi_5 - \cancel{\phi_3} = \phi_5 - \phi_1   \\
      \Phi_{2}+\Phi_{4}+\Phi_{6} =& \cancel{\phi_5} - \phi_1  + \phi_7 - \cancel{ \phi_5} = \phi_7 - \phi_1
      \end{split}
\end{equation}
The $q^{th}$ harmonic phase is equal to:
\begin{equation}\label{eqn:SB_phi}
    \phi_{q}=\phi_{1}+\sum_{m\in\text{even}}^{m=q-1}\Phi_{m}
\end{equation}
The visibility of the SB is determined by:
\begin{equation}\label{eq:visibility}
    V=\frac{2c_{q}c_{q+2}}{c_{q}^{2}+c_{q+2}^{2}}
\end{equation}
The calculated visibility referred to in Fig.4 (b-c) is calculated by the above expression, where the values of the coefficients $c_q$ are obtained from the harmonic spectra. 
To determine the visibility of the RABBIT interferogram i.e. the 'observed' data, we fitted a cosine function with a DC offset to each SB and calculated the ration between the oscillation amplitude with the DC amplitude, in accordance with the definition in eq. \ref{eq:visibility}. 

\section{Complementary results}
\subsection{1D model HHG dependence on barrier width}

The lifetime of the MS increases with both the barrier width and height  \cite{zavin_one-dimensional_2004,nussenzveig_poles_1959}. As the barrier becomes wider, the MS wavefunction becomes more localized within the potential well, increasing the lifetime. To examine how this influences rHHG, we repeated our simulation, sweeping over width ranging from 1.5 to 2.5 a.u. This corresponds to lifetime values ranging from $0.072T$ to $0.25T$.
\begin{figure}[htbp]
  \centering
  \includegraphics[width=0.8\textwidth]{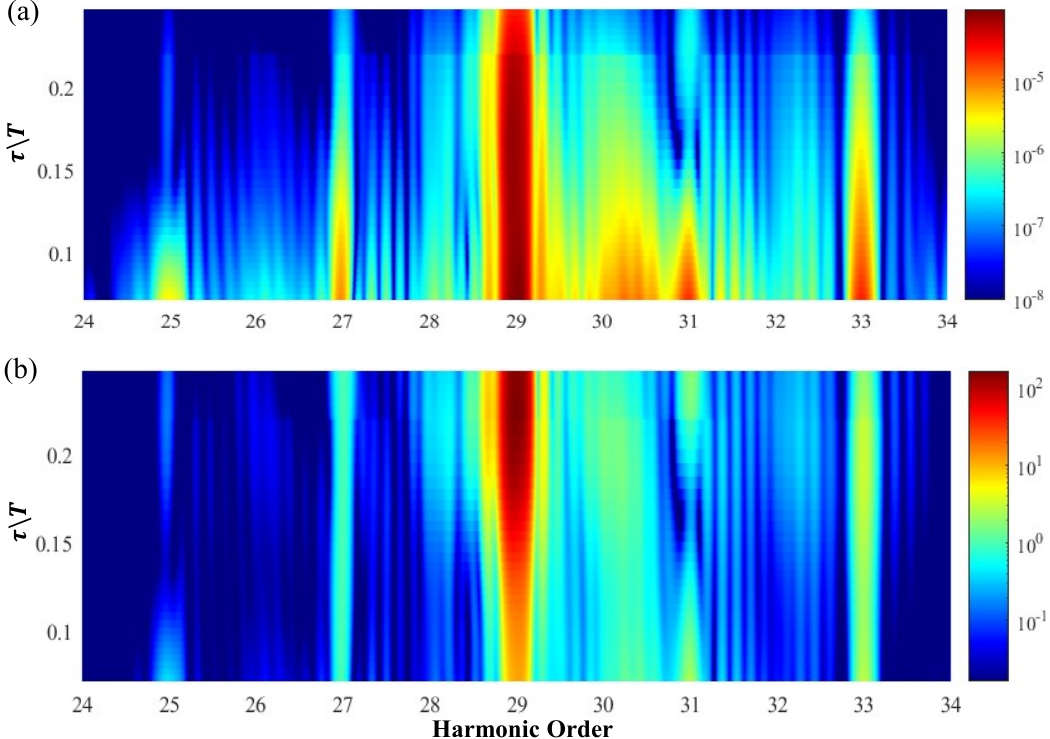}
  \caption{Harmonic spectra for varying MS lifetimes. The lifetimes are normalized by the driving period. \textbf{(a)} A clear resonance enhancement is observed at H29. The harmonic yield decreases across all harmonic orders as the MS lifetime increases, though this decrease is less pronounced for H29. \textbf{(b)} The same spectra, normalized to the intensity of H27. The intensity of H29 represents the enhancement factor, which increases with MS lifetime.}\label{fig:Barrier_spectra}
\end{figure}
We present the spectra obtained from these simulations in Fig. \ref{fig:Barrier_spectra} (a). A clear resonance enhancement is observed at H29. The yield of non-resonant harmonics generally decreases as the lifetime increases. This trend is consistent with a recollision-based model as discussed in the main text. Although the yield of the resonant harmonic also decreases, it does so to a lesser extent. In \ref{fig:Barrier_spectra} (b)  the same spectra are shown but normalized to the intensity of H27, such that the intensity of H29 reflects the enhancement factor ($I_{29} / I_{27}$). As the lifetime increases, the enhancement factor also increases.

\subsection{1D model HHG dependence on PML position}
\begin{figure}[ht]
  \centering
  \includegraphics[width=0.8\textwidth]{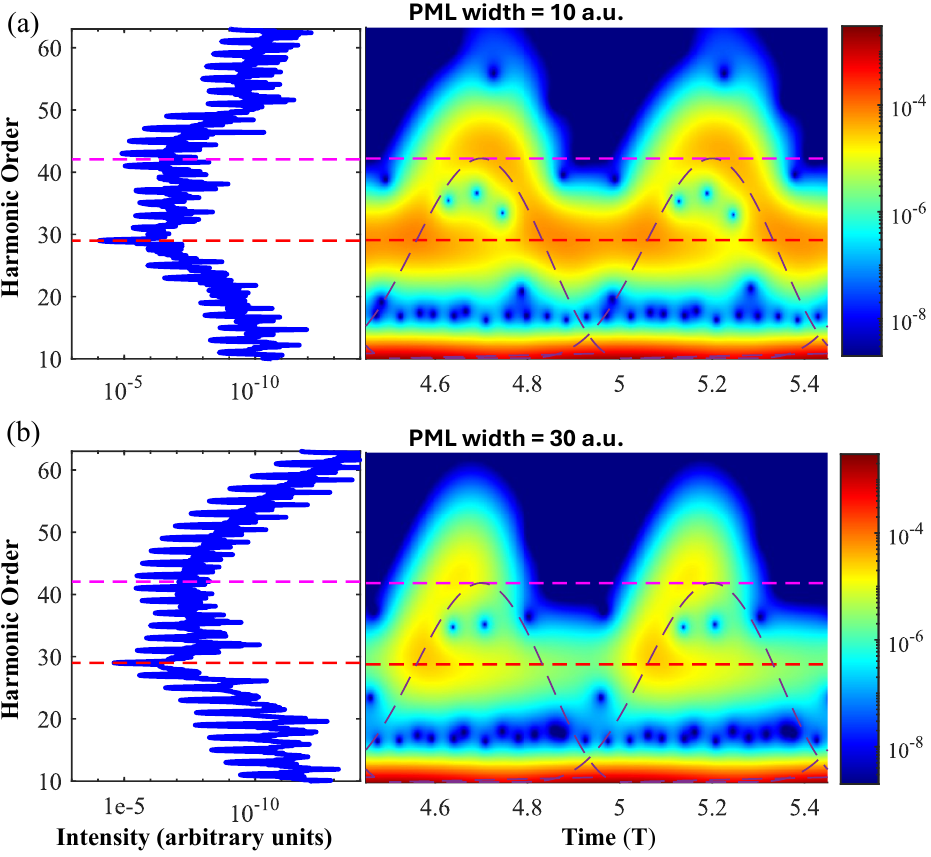}
  \caption{Gabor transform (right) and corresponding spectra (left) for different PML width. The PML is positioned at the boundary of the simulation domain, and extends inwards towards the center. \textbf{(a)} PML width of 10 a.u. Both long and short trajectories are clearly visible in the Gabor transform. \textbf{(b)} PML width of 30 a.u. The long trajectories are suppressed, diminishing the intensities of the highest harmonics, including H29.}\label{fig:pml_spectra}
\end{figure}

\begin{figure}[ht]
  \centering
  \includegraphics[width=\textwidth, trim={0 8.2cm 2.8cm 0 0},clip]{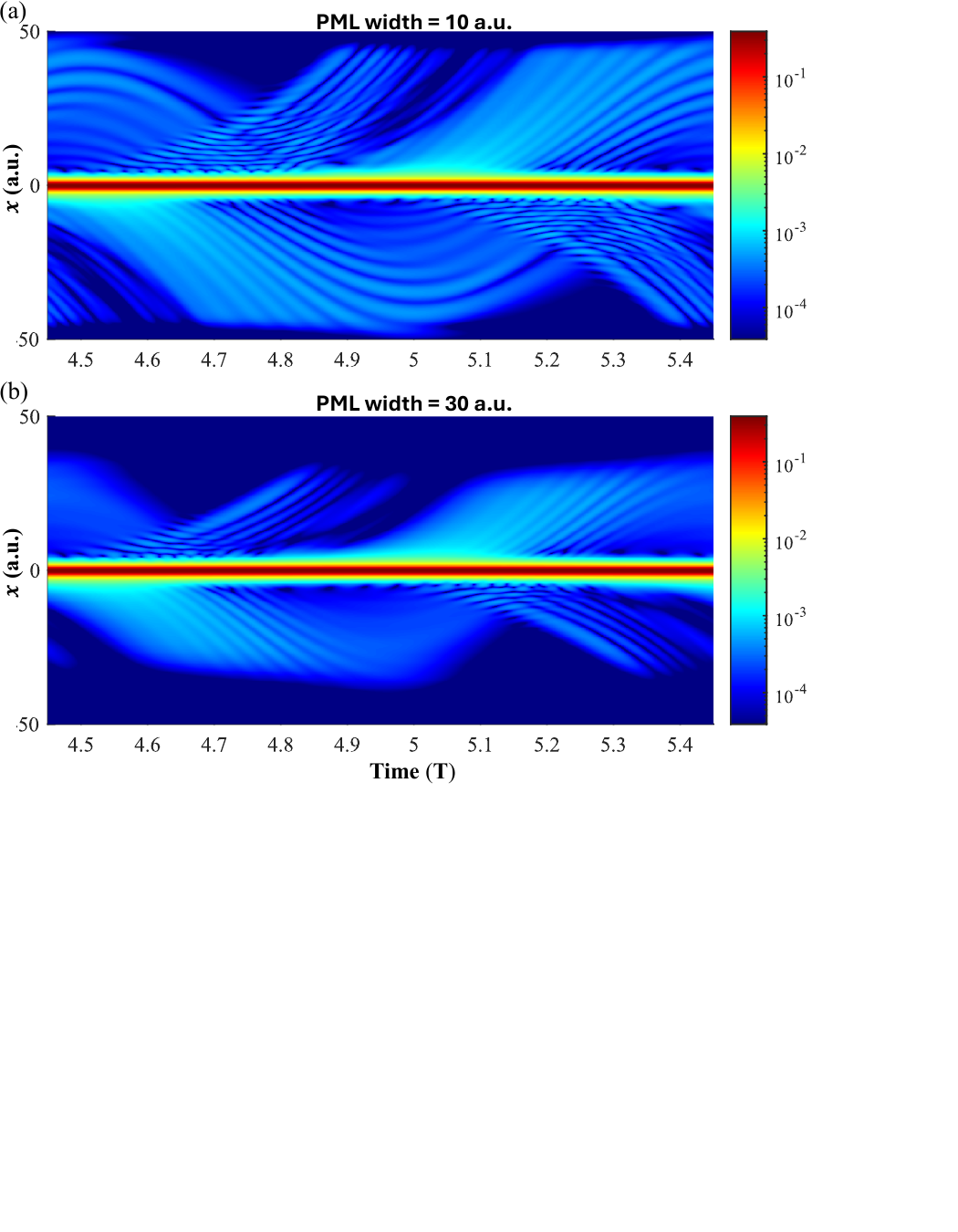}
  \caption{Plot of  $\left|  \psi \left( x,t \right) \right|^2$ for different values of PML width. The PML is positioned at the box boundary, and extends inward.  \textbf{(a)} PML width of 10 a.u. Absorption of the wavefunction occurs at the boundaries of the simulation box, where only open trajectories of non-returning electrons are absorbed. \textbf{(b)} PML width of 30 a.u. The long trajectories are absorbed, preventing recombination}\label{fig:WF_spectra}
\end{figure}

In our 1D simulation, we employed a PML (Perfectly Matched Layer) at the boundaries of the simulation domain to suppress reflections. The PML also served as a diagnostic tool to test whether H29 is emitted through tunnel ionization followed by recombination. By extending the absorbing region inward from the boundaries, we could selectively suppress the long trajectories associated with HHG. Fig. \ref{fig:pml_spectra} shows the Gabor transform and corresponding spectra for two different PML widths. In the top panel (a), the PML extends 10 a.u. into the simulation domain, remaining well outside the region reached by the calculated electron trajectories corresponding to the cutoff energy. The bottom panel (b) presents the results with a PML width of 30 a.u. which extends well beyond the range of the long trajectories associated with H29. As expected, these trajectories are clearly suppressed in the Gabor transform and the corresponding harmonics. Most notably, H29 exhibits a marked decrease in intensity in the spectrum.
The suppression of the long trajectories is also evident by examining the wavefunction directly, as shown in Fig. \ref{fig:WF_spectra}, where $\left|  \psi \left( x,t \right) \right|^2$ is plotted. In the top panel (a) , with a PML width of 10 a.u., the absorption of the wavefunction is clearly visible at the boundaries of the box, where only open trajectories of non-returning electrons are absorbed. In the bottom panel (b), where the PML extends 30 a.u. units inward, the long trajectories are absorbed, preventing recombination. 

\subsection{TDDFT absorption/photoemission calculations}

\begin{figure}[htbp]
  \centering
  \includegraphics[width=\textwidth]{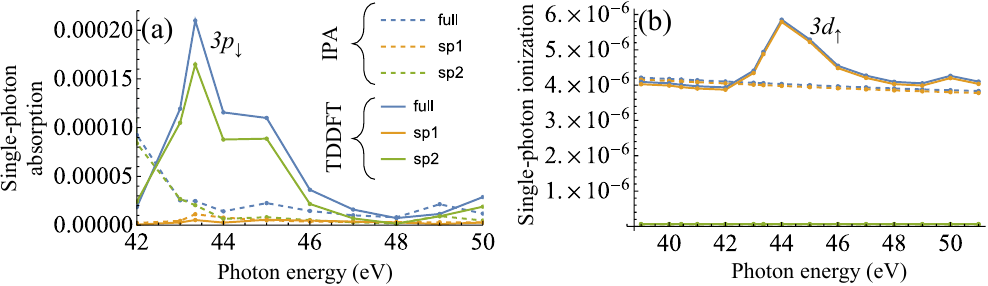}
  \caption{Single-photon absorption and ionization spectra in Cr$^+$ calculated within TDDFT and the IPA around the resonance, seperated to spin channel contributions}\label{fig:single_photon}
\end{figure}

We further present here complementary absorption and photoionization cross-section calculations performed within TDDFT and the IPA. Fig. \ref{fig:single_photon} presents the key results, showing that single-photon absorption at the resonance wavelength is dominated by the spin-down channel (i.e. $3p$ down states), which only arises when electronic correlations are considered. In other words, in the IPA there is no resonant absorption peak at the resonance energy, preventing a strong transition of $3p$ levels into the AIS (the overall enhancement in absorption with correlations included in the simulations is $\sim$ 1000$\%$). Contrarily, photoionization cross section at the resonance are dominated by the spin-up electron channel (i.e. $3d$ up states), as expected from their lower ionization potential. Notably, there is also a substantial enhancement of photoemission from the $3d$ up states if correlations are included in the simulation (about 50$\%$ enhancement), though not resonantly a very sharp enhancement as expected if resonance states are involved. This could indicate a more intricate channel by which the $3d$ photoionization enhancement is still mediated by the resonance through interaction with the $3p$ down electrons, supporting our interpretation in the main text for the enhancement mechanism. 

\clearpage
\bibliography{references}